\documentstyle[seceq,preprint,epsf]{jpsj}

\newcommand{\RJed}{J_{\mbox{\small ed}}}
\newcommand{\RJep}{J_{\mbox{\small ep}}}
\newcommand{\RJcd}{J_{\mbox{\small cd}}}
\newcommand{\RJcp}{J_{\mbox{\small cp}}}
\newcommand{\RJR}{J_{\mbox{\scriptsize R}}}

\newcommand{\RSG}{\Delta_{\mbox{\small s}}}
\newcommand{\RTp}{T_{\mbox{\small p}}}
\newcommand{\ReT}{\epsilon_{\mbox{\scriptsize T}}}

\newcommand{\Rdxy}{d_{xy}}
\newcommand{\Rdxz}{d_{xz}}
\newcommand{\Rdyz}{d_{yz}}
\newcommand{\Rpz}{p_{z}}

\newcommand{\RNA}{N_{\mbox{\scriptsize A}}}
\newcommand{\RmB}{\mu_{\mbox{\scriptsize B}}}
\newcommand{\RkB}{k_{\mbox{\scriptsize B}}}

\newcommand{\Rbk}{\mbox{\boldmath{$k$}}}
\newcommand{\Rbq}{\mbox{\boldmath{$q$}}}

\def\sf{\rm}

\def\runtitle{Orbital Order Effect for CaV$_4$O$_9$}
\def\runauthor{Nobuyuki {\sc Katoh} and Masatoshi {\sc Imada}}

\title
{
Orbital Order Effect of Two-Dimensional Spin Gap System for CaV$_4$O$_9$
}

\author
{ 
Nobuyuki {\sc Katoh}
\footnote{Present address: 
National Research Institute for Metals,
1-2-1, Sengen, Tsukuba 305, Ibaraki
}
and Masatoshi {\sc Imada}
\footnote{To be published in J.Phys.Soc.Jpn. Vol.67 No.2 (1998)}
}

\inst
{
Institute for Solid State Physics, University of Tokyo, 
7-22-1, Roppongi, Minato-ku, Tokyo 106\\
}

\recdate
{
April 15,1997
}

\abst
{
Effects of possible orbital order in magnetic properties of 
two-dimensional spin gap system for CaV$_4$O$_9$ 
are investigated theoretically. 
After analyzing experimental data, 
we show that single orbital models assumed in the literature 
are insufficient to reproduce the data. 
To understand the origin of the discrepancy, 
we assume that 
in $d^1$ state of V, 
$\Rdxz$ and $\Rdyz$ orbitals have 
substantial contributions in 
the lowest-energy atomic level which leads to a double-degeneracy. 
We study possible configurations of the orbital order. 
By exact diagonalization and perturbation expansion, 
we calculate the susceptibility, 
wavenumber dependence of low-lying excitations 
and equal-time spin-spin correlations which is related 
to integrated intensity of the neutron inelastic scattering. 
These quantities sensitively depend on the configuration of the orbital order. 
The calculated results for some configurations of the 
orbital order reproduce many experimental results 
much better than the previous single-orbital models. 
However some discrepancy still remains to  
completely reproduce all of the reported experimental results. 
To understand the origin of these discrepancies, 
we point out the possible importance of the partially occupied 
$\Rdxy$ orbital in addition to orbital order of partially filled 
$\Rdxz$ and $\Rdyz$ orbitals.
}

\kword
{
CaV$_4$O$_9$, spin gap, orbital order
}

\begin{document}
\sloppy
\maketitle

\section{Introduction}

Since the spin gap was discovered in a typical two-dimensional material 
CaV$_4$O$_9$,\cite{STaniguchi-JPSJ64-2758} 
the mechanism of the spin gap formation in this system  has been intensively 
studied by many theoretical methods.
\cite{NKatoh-JPSJ64-4105,KUeda-PRL76-1932,MPGelfand-PRL77-2794,KSano-JPSJ65-1514,KTakano-preprint,MAlbrecht-PRB53-2945,OAStarykh-preprint,MTroyer-PRL76-3822,SRWhite-PRL77-3633,TMiyazaki-JPSJ65-2370}

This material consists of VO$_5$ pyramid layers. 
Crystal structure of the VO$_5$ pyramid layer is shown in 
Fig.\ref{fig-lattice-structure}(a).
The oxygens constitute a complete square lattice 
while in the square lattice of the vanadium atoms, 
1/5 of them are depleted. 
The vanadium atom is nearly located at the center of the pyramid 
constructed from four oxygens in the layer and 
one apical oxygen. 
Each VO$_5$ pyramid is connected by edge sharing.  
The unit cell of the layer includes two edge-shared plaquettes of 
V atoms shown in 
Fig.\ref{fig-lattice-structure}(a),  
because the apical oxygen of one edge-shared plaquette 
is located above the VO$_2$ plane 
while that of the other one is below the plane.

\begin{figure}
%%%\hfil
%%%\epsfile{file=Fig1.ps,scale=0.5}
%%%\hfil
\caption{
(a) Crystal structure of VO$_5$ pyramid layer.
Full circles represent vanadium atoms, 
while open circles are oxygen atoms. 
Apical oxygens are omitted in this figure. 
Shaded square area represents the unit cell of this system. 
Vectors $\mbox{\boldmath $a$}_1$ and $\mbox{\boldmath $b$}_1$ are
the unit lattice vectors ; 
$\mbox{\boldmath $a$}_1=3\mbox{\boldmath $x$}-\mbox{\boldmath $y$}$ and 
$\mbox{\boldmath $b$}_1=\mbox{\boldmath $x$}+3\mbox{\boldmath $y$}$, 
respectively. 
Here $\mbox{\boldmath $x$}$ and $\mbox{\boldmath $y$}$ 
represent the unit lattice vectors of the square lattice composed of oxygens 
defined as 
$\mbox{\boldmath $x$}=a\mbox{\boldmath $e$}_x$
and 
$\mbox{\boldmath $y$}=a\mbox{\boldmath $e$}_y$. 
Parameter $a$ is a lattice constant while 
$\mbox{\boldmath $e$}_x$ and $\mbox{\boldmath $e$}_y$ 
are unit vectors.
(b) Four kinds of the superexchange couplings. 
Bold-gray, bold-black, thin-gray and thin-black lines 
represent the spin exchange couplings,
$J_{\mbox{\scriptsize ep}}$,
$J_{\mbox{\scriptsize ed}}$,
$J_{\mbox{\scriptsize cp}}$ and 
$J_{\mbox{\scriptsize cd}}$, respectively.
}
\label{fig-lattice-structure}
\end{figure}

Since the valence of V atom is 4+, the $d$ electron on V atom can be 
treated as a nearly localized spin with $S=1/2$. 
In the literature, 
it has been assumed that a non-degenerate orbital 
on V atom is occupied. 
\cite{NKatoh-JPSJ64-4105,KUeda-PRL76-1932,MPGelfand-PRL77-2794,KSano-JPSJ65-1514,KTakano-preprint,MAlbrecht-PRB53-2945,OAStarykh-preprint,MTroyer-PRL76-3822,SRWhite-PRL77-3633,TMiyazaki-JPSJ65-2370}
Based on this assumption, the $S=1/2$ antiferromagnetic 
Heisenberg (AFH) model with the nearest-neighbor 
and the next-nearest-neighbor exchange couplings has been introduced. 
When the nearest-neighbor exchange couplings are only 
taken into account, the system is described by 
the AFH model on the square lattice of the plaquettes, 
where each plaquette consists of four V atoms 
almost centered in the edge-shared pyramid. 
(We call this lattice plaquette lattice.) 
On the other hand, 
when only the next-nearest-neighbor exchange couplings are considered, 
the system is described by 
two disconnected square lattice of the plaquettes,  
where each plaquette is twice as large as and 45$^{\circ}$ tilted 
from the plaquette element of the previous case. 
In this case each plaquette consists of four corner-shared pyramid. 
Here, four different kinds of spin exchange couplings are introduced, 
as is shown in 
Fig.\ref{fig-lattice-structure}(b). 
The spin exchange couplings between V atoms in the edge-shared 
and the corner-shared plaquettes are represented by 
$\RJep$ and $\RJcp$, respectively.  
While the spin exchange couplings connecting the 
edge-shared and the corner-shared plaquettes are 
$\RJed$ and $\RJcd$, respectively. 
In the literature, the case where $\RJed \simeq \RJep$ is taken as $J$ 
and $\RJcd \simeq \RJcp$ is $J'$,  has been mainly studied 
theoretically since only two kinds of spin exchange couplings 
should be derived from the previous single-orbital assumptions. 
The effects of the Jahn-Teller distortion, tilting of the pyramids and 
the spin-orbit couplings may cause the differences  
between $\RJed$ and $\RJep$ or 
between $\RJcd$ and $\RJcp$. 
The theoretical methods such as 
perturbation expansions, 
\cite{NKatoh-JPSJ64-4105,KUeda-PRL76-1932,MPGelfand-PRL77-2794}
exact diagonalization, 
\cite{KSano-JPSJ65-1514}
mean field approximations, 
\cite{MAlbrecht-PRB53-2945,OAStarykh-preprint}
quantum Monte Carlo method, 
\cite{NKatoh-JPSJ64-4105,MTroyer-PRL76-3822}
high-temperature expansions
\cite{MPGelfand-PRL77-2794,KSano-JPSJ65-1514,KTakano-preprint}
and the density matrix renormalization group method 
\cite{SRWhite-PRL77-3633}
have led to the result that 
this system has a spin gap in the region around 
$\RJep > \RJed$ without frustration $(J'=0)$ 
or in the region $J'\simeq 0.5J$. 
Here the spin gap is defined as the energy difference 
between the singlet ground state and the lowest triplet states. 
In the above mentioned regions, 
the origin of the spin gap is ascribed to the edge-shared 
plaquette singlet. 
The energy dispersion of the triplet states 
has been calculated by the perturbation expansion 
from the edge-shared plaquette singlet.
\cite{NKatoh-JPSJ64-4105,KUeda-PRL76-1932,MPGelfand-PRL77-2794}
The wavenumber of the lowest energy excitation is ($\pi$,$\pi$) 
in the absence of the frustration,  
which is also obtained by the variational Monte Carlo method. 
\cite{TMiyazaki-JPSJ65-2370}
However, it shifts to incommensurate wavenumbers 
as the frustration due to $J'$ becomes relatively large.

Recently, the neutron inelastic scattering study for single crystal 
of CaV$_4$O$_9$ has been performed. 
\cite{KKodama-JPSJ65-1941,KKodama-preprint}
The experiments indicate that 
the predictions from the theoretical studies  
for the conventional model are inconsistent with 
the experimental results. 
First of all, the wavenumber of the lowest energy excitation 
given by the experiments is (0,0) in the magnetic first Brillouin zone, 
which is different from the theoretical results. 
Here, the magnetic first Brillouin zone in the wavenumber space 
is expanded by the vectors 
$\tilde{\mbox{\boldmath $a$}_2}=\tilde{\mbox{\boldmath $a$}_1}
+\tilde{\mbox{\boldmath $b$}_1}$  
and 
$\tilde{\mbox{\boldmath $b$}_2}=-\tilde{\mbox{\boldmath $a$}_1}
+\tilde{\mbox{\boldmath $b$}_1}$, 
as shown in Fig.\ref{fig-zone}. 
It is $\sqrt{2}\times\sqrt{2}$ times larger than the first Brillouin zone 
of the unit cell expanded by the reciprocal lattice vectors 
$\tilde{\mbox{\boldmath $a$}_1}=\mbox{\boldmath $a$}_1/10a^2$ 
and 
$\tilde{\mbox{\boldmath $b$}_1}=\mbox{\boldmath $b$}_1/10a^2$, 
where $a$ is a distance between the nearest-neighbor V atoms.  
Secondly, if the spin exchange couplings between V atoms 
are determined so as to reproduce the dispersion of 
the lowest triplet excitations in the experimental results, 
a large difference is resulted between the couplings, i.e. 
$\RJcd =0.088 \RJcp$ and 
$\RJep = \RJed =0.395 \RJcp$, 
\cite{KKodama-preprint}
which is hard to understand within the conventional framework.

\begin{figure}
%%%\hfil
%%%\epsfile{file=Fig2.ps,scale=0.5}
%%%\hfil
\caption{
Brillouin zone of two-dimensional VO$_2$ plane in CaV$_4$O$_9$ lattice. 
Square surrounded by broken lines is the first Brillouin zone of the lattice, 
while the square with bold-solid lines is the magnetic first Brillouin 
zone obtained by the experiment. 
\cite{KKodama-preprint} 
The wavenumber $\Rbq$ is represented by the vectors 
$\tilde{\mbox{\boldmath $a$}}_1$ and 
$\tilde{\mbox{\boldmath $b$}}_1$. 
While 
$\Rbk$ is described by 
$\tilde{\mbox{\boldmath $a$}_2}=\tilde{\mbox{\boldmath $a$}_1}
+\tilde{\mbox{\boldmath $b$}_1}$ and 
$\tilde{\mbox{\boldmath $b$}_2}=-\tilde{\mbox{\boldmath $a$}_1}
+\tilde{\mbox{\boldmath $b$}_1}$, respectively. 
Here, the lattice constant between the nearest-neighbor  V atoms is $a$.
}
\label{fig-zone}
\end{figure}

In order to understand the contradiction between the experimental results  
and the theoretical ones, 
we should reconsider the mechanism of the spin gap formation 
for CaV$_4$O$_9$. 
In this paper, we study effects of orbital degeneracy and 
orbital order which has been neglected in the literature. 
The crystal field from the oxygen ions on the corners of the pyramid 
lifts the degeneracy of $t_{2g}$ orbitals in the atomic level. 
However, two orbitals whose wavefunctions are expanded by 
$\Rdxz$ and $\Rdyz$ orbitals may still be degenerate  
even in this crystal field. 
Then the effective Hamiltonian sensitively depends on 
configuration of the occupied $d$ orbitals 
through the mechanism of the superexchange interaction 
if the occupied orbitals have substantial contributions from the 
$\Rdxz$ and $\Rdyz$ orbitals. 
In Section 2, we introduce several possible effective spin Hamiltonians 
with the orbital order. 
It is found that strength of spin exchange couplings strongly 
depends on patterns of the orbital occupancy. 
This may explain the appearance of large difference in the 
spin exchange couplings needed to reproduce experimental results. 
In Section 3, we estimate the strength of the spin exchange couplings from 
analyses of the temperature dependence of the 
uniform magnetic susceptibility. 
We use the exact diagonalization (ED) method. 
Although the system size tractable by the ED is limited, 
the cases we studied show rather small system size dependence 
due to the opening the spin gap which makes it possible to 
infer the thermodynamic limit. 
Then we calculate the energy dispersion of the triplet states and 
the wavenumber dependence of the equal-time spin-spin correlation 
which is related to the integrated intensity of the neutron inelastic 
scattering within the perturbation expansion (PE).   
The important elements to explain the experimental results 
are pointed out.  
The origin of the spin gap in each case of the orbital order 
is also discussed. 
In Section 4, we compare these theoretical results with 
the experimental ones and discuss the importance of the orbital order. 
Section 5 is devoted to summary.

\section{Effective Hamiltonians}

For the purpose of understanding the problems mentioned in 
the previous section, 
we consider several models for the spin gap formation. 
The Hamiltonian is given by the AFH model written as 
\begin{equation}
   H = \sum_{<i,j>} J_{ij} 
          \mbox{\boldmath $ S $}_{i} 
          \mbox{\boldmath $ \cdot S $}_{j}, 
\label{Hami}
\end{equation}
where $\mbox{\boldmath $ S $}_{i}$ represents the 
spin operator with $S=1/2$ at $i$ site 
and $J_{ij}$ is the spin exchange coupling between 
the spins at $i$ and $j$ sites.

In this study, 
we assume that the orbitals giving the lowest energy in the atomic level 
due to the crystal field are doubly degenerate. 
In other words, the ground state wavefunction has 
substantial weight of the $\Rdxz$ and $\Rdyz$ orbitals 
and that the $d$ electron occupies these orbitals at least partially.  
Below we refer to occupied orbitals simply as 
$\Rdxz$ or $\Rdyz$ orbitals in this paper. 
However, it does not necessarily mean that 
those consist of pure $\Rdxz$ and $\Rdyz$ orbitals. 
What we need to make the proposals in this paper relevant is 
the double-degeneracy of occupied orbitals with 
$\Rdxz$ and $\Rdyz$ component contained. 
Even small contribution of 
$\Rdxz$ and $\Rdyz$ orbitals may make the exchange coupling 
sensitively dependent on the occupied orbitals. 
Although the static Jahn-Teller distortion or tilting 
of the pyramid could lift the degeneracy of 
$\Rdxz$ and $\Rdyz$ orbitals, 
we do not consider this effect within this study, 
because such static distortions have not been observed so far.  
Therefore, we investigate effects of the orbital order for 
the $\Rdxz$ and $\Rdyz$ orbitals. 
As is mentioned later, the strength of the spin exchange couplings 
between V atoms depends on the configuration of the occupied $d$ orbitals 
on V atoms.

We consider the spin exchange couplings 
between the nearest as well as between 
the next-nearest neighbor V atoms 
through the superexchange mechanism, 
as is shown in  
Fig.\ref{fig-orbital-exchange}. 
%%%
%%%
\begin{figure}
%%%\hfil
%%%\epsfile{file=Fig3.ps,scale=0.5}
%%%\hfil
\caption{
Lattice structure of V-O-V and configuration of occupied orbital. 
Figures (a) and (b) show the cases with the right angle of V-O-V. 
Figure (a) shows the case that the occupied $d$ orbitals on 
V atoms are different, 
while (b) is the case with the same occupied orbital. 
Figures (c) and (d) show the cases with the 
straight line of V-O-V. 
The configurations of the occupied $d$ orbitals in Fig.(c) and (d) 
are the same as those in Fig.(a) and (b), respectively. 
 }
\label{fig-orbital-exchange}
\end{figure}
%%%
%%%
In the first case, the V-O-V bonds make approximately  right angle. 
Then the superexchange coupling works relevantly through the $\Rpz$ 
orbital of the oxygen when the $d$ electron occupies $\Rdxz$ orbital 
on V atom and the other $d$ electron is localized at $\Rdyz$ orbital 
on the other V atom shown in 
Fig.\ref{fig-orbital-exchange}(a). 
However, the superexchange coupling may be 
relatively small when the both $d$ electrons occupy the same 
orbitals on each V atom shown in 
Fig.\ref{fig-orbital-exchange}(b). 
The reason is that one of the transfer integrals between $\Rpz$ orbital on 
oxygen and one of the $d$ orbitals is relatively 
small due to the symmetry of the wavefunction of these orbitals. 
In the second case, the V-O-V bonds make approximately straight line. 
Then the superexchange coupling works effectively when the $d$ electrons 
are localized at the same orbitals which extend to the oxygen as shown in 
Fig.\ref{fig-orbital-exchange}(d). 
While the superexchange coupling should be relatively 
small in the case shown in  
Fig.\ref{fig-orbital-exchange}(c). 
When the $d$ electrons occupy the orbitals 
both of which do not extend to the oxygen located between the V atoms, 
the strength of the superexchange couplings should be also relatively small.  
Consequently, the anisotropy of the spin exchange couplings arises 
through this mechanism.

Here, we introduce five models with the possible patterns of the 
orbital order as shown in Figs.\ref{fig-pattern}. 
The orbital patterns in Figs.\ref{fig-pattern}(a) and 
(b) are the cases 
where the electrons occupy the $d$ orbitals alternatingly 
within the edge-shared plaquettes 
but in a uniform way from plaquette to plaquette. 
The relevant spin exchange couplings in Fig.\ref{fig-pattern}(a) 
are the edge-shared plaquette bonds 
$\RJep$, 
while the ones in Fig.\ref{fig-pattern}(b) 
are $\RJep$ and $\RJcd$.
Figure \ref{fig-pattern}(c) shows the case 
where the plaquette unit in 
Fig.\ref{fig-pattern}(a) and (b) are placed alternatingly.  
Figure \ref{fig-pattern}(d) shows the case 
where all the occupied orbitals have the same symmetry.
Figure \ref{fig-pattern}(e) shows the case where  
the occupied $d$ orbitals  are alternating in the $x$-direction 
while are uniform in the $y$-direction.  
From here, we take the strengths of the relatively small 
spin exchange couplings as the same values. 
In reality, these value are not necessarily the same 
because these small couplings appear due to some small distortion 
of the lattice from the perfect square lattice. 
Then to study difference between them,  
these small effects have to be considered.
This point is discussed in detail in Section 4.    
In other possible patterns of the orbital order, 
the magnetic unit cell becomes more complicated. 
In this paper, 
we do not consider these more complicated possibilities.

\begin{figure}
%%%\hfil
%%%\epsfile{file=Fig4.ps,scale=0.5}
%%%\hfil
\caption{
Five possible patterns of orbital order 
viewed in the projection to the $xy$ plane.  
Bold lines connecting V atoms 
represent the dominant superexchange couplings.   
The symbols on V atoms represent the 
$\Rdxz$ and $\Rdyz$ orbitals schematically.  
(a) A case with the alternating pattern 
of the orbitals in the edge-shared plaquette.
The filled orbitals are ``tiled'' 
in a uniformed way among different plaquettes.
(b) This is the same as the one in Fig.(a).  
However, the $\Rdxz$ and $\Rdyz$ orbitals are 
replaced each other. 
(c) The case where the orbital in all the nearest-neighbor 
V atoms are occupied always in the alternating configuration. 
(d) The case that all the orbitals are occupied uniformly.
(e) The case that the occupied orbitals are 
alternating in the $x$-direction 
while are uniform in the $y$-direction.
 }
\label{fig-pattern}
\end{figure}

\section{Results}

In this section, we study magnetic properties 
for the models in Figs.\ref{fig-pattern}(a), (b) and (c) 
in more detail than for (d) and (e) 
since the configuration of the occupied orbitals 
in the edge-shared plaquette
for these models has four-fold rotational symmetry 
which appears to be consistent with experimental observations. 
\cite{KKodama-preprint}

\subsection{Uniform Magnetic Susceptibility}

In this subsection, we determine the amplitude of the relevant 
spin exchange couplings and the other relatively small ones 
to reproduce the temperature dependence of 
the uniform magnetic susceptibility $\chi$ given by the experiments. 
Here, 
we assume that 
eq.(\ref{Hami}) is 
an effective spin Hamiltonian of 
magnetic properties below 700K.  
We calculate the $\chi$ of 16-site system 
by the ED with the periodic boundary condition. 
The uniform magnetic susceptibility is defined as
\begin{equation}
 \chi=\frac{\beta}{N}\frac{\mbox{Tr} \sum_i \sum_j
  S^{z}_iS^{z}_j
 \exp({-\beta \epsilon_n})}
{\mbox{Tr}\exp({-\beta \epsilon_n})}, 
\label{eq-chi}
\end{equation}
where $\beta$ represents the inverse temperature 
and $N$ is the number of site. 
In terms of comparison with the experimental data in the unit of emu/g, 
we should multiply our data with energy scale $J$ by a factor 
$4 \RNA g^2 \RmB /J \RkB M$ 
with 
$\RNA$ the Avogadro number, 
$\RmB$ the Bohr magneton, 
$\RkB$ the Boltzmann constant and 
$M$ the gram per mole. 
The factor 4 comes from the number of V atoms in a unit cell.
The parameter $J$ depends on each model. 
Here, the $g$-factor is taken as a fitting parameter. 
The experimental data show a peak at a temperature about $\RTp \simeq 110$K.
The temperature $T^*$ where the amplitude of $\chi$ becomes a half of that 
at $\RTp$ is about 595K. 
We choose the amplitude of the spin exchange couplings 
so as to give the best fit between the ED and the experimental results 
in the region above $\RTp$.

Figure~\ref{fig-chi}(a) shows the temperature dependence of the $\chi$ 
for the model shown in Fig.\ref{fig-pattern}(a). 
The fitting of the data leads to the spin exchange coupling $\RJep$ is 183K, 
while relatively small ones are  
$\RJed = \RJcp = \RJcd =$97K, respectively. 
The ratio $\RJed / \RJep $ is 0.53. 
The $g$-factor is 1.71, which is rather small compared to  
$g\simeq 2.0$. 
In principle, the mechanism which reduces the value of the $g$ factor 
estimated from $\chi$ in experiments could exist. 
One possibility is an effect of small spin-orbit couplings. 
However, we do not discuss this point in this paper.   
As a reference, the data not only for 16-site system 
but also for 8-site system are shown in Fig.\ref{fig-chi}(a). 
The size dependence appears below $\RTp$. 
However, the finite size effect is small above $\RTp$, 
which justifies the present estimate of the spin exchange couplings 
to take as the value in the thermodynamic limit.

The spin gap for 8-site, 16-site and 24-site systems were calculated. 
The size dependence of the spin gap is shown in the inset in 
Fig.\ref{fig-chi}(a). 
If we assume that the fitting function for two-dimensional spin gap system 
has a form as $\RSG (N)= \RSG (\infty)+\frac{A}{N}$, 
the spin gap extrapolated to the thermodynamic limit 
is estimated to be about 
101K, namely 
$\RSG / \RJep =0.550$.  
The reason for taking this fitting function 
is as follows. The energy dispersion of the triplet excitation 
near the wavenumber with the lowest energy excitation 
is approximated as 
$\ReT (\Rbk)=
\RSG +c_x k_x^2 + c_y k_y^2$ 
for sufficiently small $k_x$ and $k_y$, 
where $\RSG$ represents the bulk spin gap. 
Using the periodic boundary conditions, 
the wavenumber $k_x(k_y)$ has a 
$L_x^{-1}(L_y^{-1})$ correction, 
where $L_x(L_y)$ is the number of sites in the 
$x(y)$-direction. 
Because $L_x$ and $L_y$ are taken as $N^{-1/2}$ in two dimensions, 
the above fitting function is obtained. 
If the energy dispersion can be approximated by the quadratic form 
in a wide range of the wavenumber space, 
the fitting function reduced from the quadratic dispersion 
may be available even in small size systems. 
This model may belong to this case. 
%%%
%%%
\begin{figure}
%%%\hfil
%%%\epsfile{file=Fig5a.ps,scale=0.5}
%%%\hfil
%%%\end{figure}
%%%\begin{figure}
%%%\hfil
%%%\epsfile{file=Fig5b.ps,scale=0.5}
%%%\hfil
%%%\end{figure}
%%%\begin{figure}
%%%\hfil
%%%\epsfile{file=Fig5c.ps,scale=0.5}
%%%\hfil
%%%\end{figure}
%%%\begin{figure}
%%%\hfil
%%%\epsfile{file=Fig5d.ps,scale=0.5}
%%%\hfil
%%%\end{figure}
%%%\begin{figure}
%%%\hfil
%%%\epsfile{file=Fig5e.ps,scale=0.5}
%%%\hfil
\caption{
Temperature dependence of uniform magnetic susceptibility for 
five models. 
Figures (a), (b), (c), (d) and (e) correspond to  
$\chi$ for the models in 
Figs.\ref{fig-pattern} (a), (b), (c), (d) and (e), respectively. 
Bold-gray lines represent the experimental result. 
Insets show the extrapolation of the spin gap as a function of $N$ 
in the ED results.
}
\label{fig-chi}
\end{figure}
%%%
%%%

For the case shown in Fig.\ref{fig-chi}(b), 
we take the relevant spin exchange couplings as 
$\RJcd >\RJep$, for example, 
$\RJep =0.9 \RJcd$. 
The reason is discussed in the next subsection. 
Figure~\ref{fig-chi}(b) shows the temperature dependence of the $\chi$. 
In a  similar way, 
the relevant spin exchange couplings are estimated to be 
$\RJcd \sim 185$K and $\RJep \sim 166$K,
while the relatively small ones are 
$\RJcp = \RJed =85$K. 
The ratio $\RJcp / \RJcd $ is 0.46. 
The g-factor is 1.72. 
In this case, 
one should be careful in estimating the spin gap from the data only 
for small size systems 
since the system seems to be near the critical point where 
the spin gap vanishes. 
Beyond the critical point 
when the spin gap disappears, the energy dispersion of the triplet 
states near the lowest excitation is expected to have a form as 
$\ReT(\Rbk)=
\sqrt{c_x k_x^2 + c_y k_y^2}$, 
namely, a linear dispersion. 
As the system with the spin gap is close to the critical point, 
the wavenumber space where the form of the energy dispersion 
is approximated to be quadratic becomes narrower and narrower  
while the linearly dispersive region becomes wider. 
Then the leading term of the finite-size correction 
may crossover from $N^{-1/2}$ for small system size to 
$N^{-1}$ for large system size. 
Here we take the form 
$\RSG (N)= \RSG (\infty)+\frac{A}{\sqrt{N}}$ 
as the fitting function. 
We have tried extrapolations by using two types of the fitting function.  
However, we found that the leading term of the finite size correction 
is $N^{-1/2}$ rather than $N^{-1}$ in this case 
due to the limitation of the system size. 
Using this fitting function, 
the spin gap extrapolated to the thermodynamic limit is 
estimated to be about 
34K, or $\RSG / \RJcd =0.186$. 
From the above reason, 
this estimate of the spin gap is expected to give a slight underestimate.  
In this paper, in order to estimate the bulk spin gap, 
these two fitting functions are 
employed according as the situation.

In Fig.\ref{fig-pattern} (c), 
the relevant spin exchange couplings are divided into two groups. 
One is the nearest-neighbor spin exchange coupling $J$ and 
the other is the next-nearest-neighbor one $J'$.  
We consider the case that $J'$ is larger than $J$ 
and the other relatively small exchange couplings are neglected. 
The reason is discussed in the next subsection. 
Then the parameters $J$ and $J'$ are estimated to be 90K and 237K, 
respectively. 
The ratio $J/J'$ is 0.38. The $g$-factor is $1.68$. 
The relatively large spin gap survives in the thermodynamic limit 
and is estimated to be 
$\RSG /J'\simeq 0.564$ or 133K 
when we use the fitting function 
$\RSG (N)= \RSG (\infty)+\frac{A}{N}$.

The values of the spin gap for the first and the third model are 
close to the one in the experimental results.  
However, in the second case, the spin gap value  
is much smaller than the experimental one.

\subsection{Origin of Spin Gap and Triplet Energy Dispersion}

We consider the mechanism of the spin gap formation. 
To investigate the dispersion of triplet excitations, 
the perturbation method is useful  
since it is found by the analysis mentioned in the previous subsection 
that the strength of the relevant spin exchange couplings 
is relatively large compared to the others.
In the case shown in Fig.\ref{fig-pattern} (a), 
the term in eq.(\ref{Hami}) 
including the relevant spin exchange coupling $\RJep$ 
is treated as the unperturbed Hamiltonian, while 
the other terms are the perturbed ones. 
The strength of the other spin exchange couplings is taken as $J$. 
The singlet ground state of the isolated edge-shared 
plaquette is the resonating valence bond (RVB) singlet. 
Then the ground state of the unperturbed Hamiltonian 
is the product state of the RVB singlets on the edge-shared plaquettes. 
The origin of the spin gap is understood by 
the edge-shared plaquette singlet. 
The lowest excited state of the plaquette is 
the extended state of the triplet pair in the RVB state. 
Then the first excited states of the unperturbed Hamiltonian 
are constructed from the triplet state on one of the 
plaquettes and the singlets on the others. 
The degeneracy is lifted by the first-order perturbation   
through the transfer of the triplet due to the translational invariance, 
which yields the dispersion of the triplet states. 
We calculate the energy difference between the ground state and 
the triplet states within the second-order PE. 
The detailed procedure is explained in Appendix A. 
Figure~\ref{fig-dispersion} (a) shows the energy dispersion of the 
triplet excitation $\RSG (\Rbk)$. 
The wavenumber shown as $(k_x,k_y)$ represents the vectors 
$k_x\tilde{\mbox{\boldmath $a$}}_2 + k_y\tilde{\mbox{\boldmath $b$}}_2$   
with 
$\tilde{\mbox{\boldmath $a$}}_2 = 
\tilde{\mbox{\boldmath $a$}_1}+\tilde{\mbox{\boldmath $b$}_1}$ and 
$\tilde{\mbox{\boldmath $b$}}_2= 
-\tilde{\mbox{\boldmath $a$}_1}+\tilde{\mbox{\boldmath $b$}_1}$. 
The first magnetic Brillouin zone is expanded by 
$\tilde{\mbox{\boldmath $a$}}_2$ and 
$\tilde{\mbox{\boldmath $b$}}_2$ as 
shown in 
Fig.\ref{fig-zone}. 
In this case, the wavenumber of the lowest excitation is (0,0).  
Since a strong antiferromagnetic correlation due to $\RJcp$ 
is larger than the one due to $\RJed$ in this model,  
the triplet states which have a strong in-phase correlation 
between the nearest-neighbor 
edge-shared plaquettes become the lowest excited states. 
The amplitude of the spin gap at $(0,0)$ calculated by the second-order PE 
is $0.575 \RJed$, which is close to the one estimated by the ED and 
is also consistent with the experimental result. 
These behaviors qualitatively reproduce the experimental results.

\begin{figure}
%%%\hfil
%%%\epsfile{file=Fig6a.ps,scale=0.5}
%%%\hfil
%%%\end{figure}
%%%\begin{figure}
%%%\hfil
%%%\epsfile{file=Fig6b.ps,scale=0.5}
%%%\hfil
%%%\end{figure}
%%%\begin{figure}
%%%\hfil
%%%\epsfile{file=Fig6c.ps,scale=0.5}
%%%\hfil
\caption{
Wavenumber dependence of triplet and singlet excited state energies  
for three models. 
Figures (a), (b) and  (c) correspond to  
the models shown in 
Figs.\ref{fig-pattern} (a), (b) and (c), respectively. 
Filled circles with error bar represent the experimental results.
\cite{KKodama-preprint}
}
\label{fig-dispersion}
\end{figure}

In the case shown in Fig.\ref{fig-pattern} (b), 
the terms with the relevant spin exchange couplings 
$\RJep$ and $\RJcd$ are treated as the 
unperturbed Hamiltonians, while the other terms are the perturbed ones. 
In the $\RJcd > \RJep$ case, 
the product state of the dimer pairs on $\RJcd$ bonds 
is the unperturbed ground state. 
Then the origin of the spin gap is ascribed to   
the corner-shared dimer singlets.  
The lowest excited states of the edge-shared plaquette 
are the product states of the triplet pair on one $\RJcd$ bond 
and the singlet pair on the other $\RJcd$ bonds. 
Because two kinds of triplet states exist, the lowest excited states 
are six-fold degenerate, which is also different from the previous case. 
We calculate the energy gap between the singlet state and 
those triplet states in a similar way. 
Figure~\ref{fig-dispersion} (b) shows the energy dispersion of the 
triplet states. 
The degeneracy is lifted by the first-order perturbation   
and the bonding and the anti-bonding triplet states appear. 
The wavenumbers of the lowest excitations are  
$(0,\pi)$ and $(\pi,0)$. 
The spin gap is $0.197 \RJcd$,  
which is consistent with the value obtained by the ED. 
In this case, the square area surrounded by the lines connecting 
the nearest-neighbor points of the lowest excitation 
becomes a half of the original first magnetic Brillouin zone 
in the wavenumber space, 
which contradicts the experimental results.

In the model shown in Fig.\ref{fig-pattern} (c), 
the ground state of the Hamiltonian given by the 
relevant spin exchange couplings is not straightforwardly obtained 
since the system has no isolated terms.  
To understand the mechanism of the spin gap formation, 
we neglect the small spin exchange couplings 
other than $J$ and $J'$ in this model.  
We consider three possible ground states 
and the origins of the spin gap for the Hamiltonian  
with only the relevant spin exchange couplings. 
To clarify three possibilities with the help of the perturbation method,  
the relevant terms are divided into two parts, 
unperturbed and perturbed terms.  
In the first case, the unperturbed Hamiltonian is taken as 
the term with $\RJep$ 
while the others are the perturbed ones. 
Then the ground state of the unperturbed Hamiltonian 
is represented by the edge-shared plaquette singlet. 
In the second case, we take the unperturbed Hamiltonian as the $\RJed$ term. 
Then the ground state is described by the edge-shared dimer singlet. 
In the third case, the unperturbed Hamiltonian is the term with 
the next-nearest-neighbor spin exchange couplings. 
Then the ground state is represented by the four-site stripe singlet. 
Since the first and the second cases have been already investigated, 
\cite{NKatoh-JPSJ64-4105,KUeda-PRL76-1932}
we concentrate on the third case. 
Then the amplitudes of the nearest and the next-nearest neighbor 
couplings are chosen as $J$ and $J'$, respectively.  
Figure~\ref{fig-gse} shows the ground state energy per site 
as a function of 
the strength of the spin exchange coupling $J/(J+J')$
calculated by the PE. 
The detailed calculation by the PE is written in Appendix B. 
%%%
%%%
\begin{figure}
%%%\hfil
%%%\epsfile{file=Fig7.ps,scale=0.5}
%%%\hfil
\caption{
Ground state energy per site 
$\epsilon_{\mbox{\small g}}/(J+J')$ 
as a function of $J/(J+J')$.
Bold, broken and dash-dotted lines represent the ground state energy 
by the second-order perturbation from the 
plaquette singlet, dimer singlet and stripe singlet, respectively. 
}
\label{fig-gse}
\end{figure}
%%%
%%%
The ground state described by  
the stripe singlet is favored when $J/(J+J')$ is smaller than 0.483,  
or the ratio $J/J'$ is smaller than 0.932.   
The system with $J=90$K and $J'=237$K estimated by the ED 
belongs to the stripe singlet phase 
as shown in Fig.\ref{fig-gse}. 
Therefore, we may conclude that the stripe singlet is a possible 
mechanism of the spin gap formation in this model. 
We calculate the triplet dispersion as shown in 
Fig.\ref{fig-dispersion}(c). 
The ground state is the product state of the stripe singlet for  
the unperturbed Hamiltonian. 
The magnetic unit cell in the real space  
contains two stripes crossing each other.  
Then the lowest triplet states in the unit cell are six-fold degenerate. 
Similarly to the analysis for the model given in Fig.\ref{fig-pattern}(b), 
the bonding and anti-bonding triplet states appear. 
The spin gap is $0.539J'$, which is close to the value 
obtained by the ED and is also close to the experimental result. 
However, the wavenumbers of the lowest excitations are 
$\pi\tilde{\mbox{\boldmath $a$}_1}$ and 
$\pi\tilde{\mbox{\boldmath $b$}_1}$ 
respectively, which is inconsistent with the experimental results. 
The magnetic periodicity is also different from the experimental one, 
since the magnetic first Brillouin zone is the same as that  
of the unit lattice.  
In this model, small exchange couplings in the corner-sharing have been 
neglected. 
They may change the quantitative feature such as band-width of the 
triplet dispersion. 
However, an effect due to them is very small 
since both of the occupied $d$ orbitals on V atoms do not extend to the oxygen.
Qualitative aspects such as the origin of the spin gap and 
the magnetic periodicity are not affected by these small exchange couplings.

\subsection{Scattering Intensity}

We investigate the scattering intensity of the neutron 
inelastic scattering given by the experiments. 
The scattering intensity is proportional to the 
Fourier component of the spin-spin correlation written as
\begin{equation}
I(\Rbq,\omega)
\propto \int dt \mbox{e}^{\mbox{i}\omega t}
\langle \mbox{\boldmath $S$}(
\Rbq,t)
\mbox{\boldmath $\cdot S$}
(-\Rbq,0) \rangle, 
\label{intensity}
\end{equation} 
where 
\begin{equation}
\mbox{\boldmath{$S$}}(
\mbox{\boldmath{$q$}},t)=N^{-\frac{1}{2}}\sum_i 
\mbox{\boldmath{$S$}}_i(t)\mbox{e}^{-\mbox{i}
\mbox{\boldmath{$q\cdot r$}}_i},
\end{equation} 
and $\langle \cdots \rangle$ represents the thermal average. 
The wavenumber $\Rbq$ are expanded by 
the vectors 
$\tilde{\mbox{\boldmath $a$}_1}$ and 
$\tilde{\mbox{\boldmath $b$}_1}$, 
namely we take $q_x$ axis in the 
$\tilde{\mbox{\boldmath $a$}_1}$ direction and 
$q_y$ in the $\tilde{\mbox{\boldmath $b$}_1}$ direction. 
When the temperature is sufficiently lower than the spin gap, 
eq.(\ref{intensity}) is approximated by the transition probability 
from the ground state $|\mbox{g.s.}\rangle$ 
to the triplet states $|n\rangle$, 
which is described as
\begin{equation}
I(\Rbq,\omega)
\propto 
\sum_{n} 
|\langle n |\mbox{\boldmath $S$}(\Rbq)
|\mbox{g.s.}\rangle|^2
\delta (\omega-E_n+E_g), 
\label{intensity-2}
\end{equation} 
where the summation is restricted to the triplet states 
and $E_n$ and $E_g$ are the energy of the triplet states and the 
ground state, respectively. 
Here, when we calculate eq.(\ref{intensity-2}), 
we use the wavefunctions of $|n\rangle$ and $|\mbox{g.s.}\rangle$ 
obtained by the PE, 
which is written in Appendix A in detail. 
The integrated intensity of the neutron inelastic scattering 
is obtained theoretically by integrating eq.(\ref{intensity-2}) 
over $\omega$, 
described as 
\begin{equation}
I(\Rbq)
\equiv  
\int d\omega I(\Rbq,\omega)
\propto 
\sum_{n} 
|\langle n |\mbox{\boldmath $S$}(\Rbq)
|\mbox{g.s.}\rangle|^2.
\label{intensity-3}
\end{equation}
Figures~\ref{fig-intensity}(a),(b) and (c) show 
the wavenumber dependence of the integrated intensity 
calculated by the PE. 
The experimental results are also shown in these figures.   
Especially, 
Figs.\ref{fig-intensity}(a) and (c) indicate important points. 
In the case shown in Fig.\ref{fig-pattern}(a), 
since the nearest-neighbor antiferromagnetic correlation due to 
$\RJep$ is dominant, the scattering intensity has a minimum 
at $(\pi/a) \mbox{\boldmath{$e$}}_x$ and becomes larger as 
the wavenumber approaches  
$(\pi/a)\mbox{\boldmath{$e$}}_x+(\pi/a)\mbox{\boldmath{$e$}}_y$. 
While in the cases shown in 
Figs.\ref{fig-pattern}(b) and (c), 
since the next-nearest-neighbor antiferromagnetic correlation is dominant, 
the scattering intensity has a maximum at $(\pi/a)\mbox{\boldmath{$e$}}_x$ 
and becomes smaller as the wavenumber approaches 
$(\pi/a)\mbox{\boldmath{$e$}}_x+(\pi/a)\mbox{\boldmath{$e$}}_y$. 
From this result, 
experimental results indicate that the 
next-nearest-neighbor antiferromagnetic correlation is strong. 
\cite{KKodama-preprint}
This supports the models in Figs.\ref{fig-pattern}(b) and (c). 
However, the wavenumber dependence for the model 
in Figs.\ref{fig-pattern}(c) is qualitatively 
different from the experimental results as shown in 
Figs.\ref{fig-intensity}(a) and (b).
Therefore, 
the experimental results are rather consistent with 
the integrated intensity for 
the model shown in Fig.\ref{fig-pattern}(b) than 
those for the models shown in Figs.\ref{fig-pattern}(a) or (c).
%%%
%%%
\begin{figure}
%%%\hfil
%%%\epsfile{file=Fig8.ps,scale=0.5}
%%%\hfil
\caption{
Wavenumber dependence of the integrated intensity for three models 
shown in Fig.\ref{fig-pattern}(a),(b) and (c).  
The wavenumber $\Rbq$ in Figures (a), (b) and (c) 
are represented by 
$\Rbq =2\pi(\tilde{\mbox{\boldmath $a$}_1}+k
\tilde{\mbox{\boldmath $b$}_1})$ with $0.5 < k < 1.5$, 
$\Rbq =2\pi(h\tilde{\mbox{\boldmath $a$}_1}+
\tilde{\mbox{\boldmath $b$}_1})$ with $1 < h < 2$ and 
$\Rbq =2\pi[(1+\delta)\tilde{\mbox{\boldmath $a$}_1}+
(1-\delta)\tilde{\mbox{\boldmath $b$}_1}]$ with $0 < \delta < 1$, 
respectively. 
Filled circles represent the experimental results  
given by Kodama {\it et al.}
\cite{KKodama-preprint}
The right ordinates represent the 
scale of the integrated intensity of the neutron inelastic scattering 
experiments. 
Here, the value 200 counts/5500kmon in the experiments corresponds 
to 0.04 obtained by the PE. 
This gives 
the best fit of the data for the model shown in 
Fig.\ref{fig-pattern}(b) with the experimental results.
%%%in considering the case shown in Fig.\ref{fig-intensity}(a), for example. 
}
\label{fig-intensity}
\end{figure}
%%%
%%%

\section{Discussion}

To understand the mechanism of spin gap in CaV$_4$O$_9$, 
we have studied the orbital-order dependence of the physical quantities 
such as the uniform magnetic susceptibility, the energy dispersion 
of the triplet states and the integrated scattering intensity. 
Before discussing these results, 
we show the temperature dependence of $\chi$ 
for the system with 
$\RJcp =171$K,  
$\RJcd =15$K and   
$\RJep =\RJed =68$K. 
These parameter values are the ones suggested by an analysis of 
the neutron inelastic scattering experiments.   
\cite{KKodama-preprint}
%%%
%%%
\begin{figure}
%%%\hfil
%%%\epsfile{file=Fig9.ps,scale=0.5}
%%%\hfil
\caption{
Temperature dependence of the uniform magnetic susceptibility 
for the model with  
$\RJcp =171$K,  
$\RJcd =15$K and   
$\RJep = \RJed =68$K. 
Broken and solid curves represent 
$\chi$ for 8-site and 16-site systems 
calculated by the ED, respectively, 
while the bold-gray line is the experimental result. 
}
\label{fig-chi-Kodama}
\end{figure}
%%%
%%%
The uniform magnetic susceptibilities for 8-site and 16-site 
systems are calculated by the ED. 
The theoretical and the experimental results are shown in 
Fig.\ref{fig-chi-Kodama}. 
The size dependence is almost negligible and 
the numerical result of the $\chi$ for 16-site system may be  
regarded as the one in the thermodynamic limit. 
Since it is difficult to estimate  
the amplitude of the spin exchange couplings and the $g$-factor 
from the high temperature expansion,  
the $g$-factor is taken as 1.7 here, which is nearly the same as the 
value obtained by the analysis mentioned in the previous section. 
Figure~\ref{fig-chi-Kodama} indicates that a qualitative difference 
between the experimental and theoretical results appears 
around the peak temperature. 
We may conclude that the parameters suggested by the analysis of 
the neutron inelastic scattering 
\cite{KKodama-preprint} 
are not consistent with the experimental result of 
the uniform magnetic susceptibility.

In Section 3, we have mainly studied the cases in 
Figs.\ref{fig-pattern}(a)-(c). 
Here, we briefly discuss the cases in Figs.\ref{fig-pattern}(d) and (e) 
for completeness.  
Figure~\ref{fig-chi}(d) and (e) show the temperature dependences of $\chi$. 
For the model in Fig.\ref{fig-pattern}(d), 
strength of the relevant spin exchange coupling $\RJR$ and 
the other small ones $J$ are estimated to be 159K and 111K, respectively. 
The ratio $J/\RJR$ is 0.70. The $g$-factor is 1.71. 
The amplitude of the spin gap is extrapolated to be about 29K 
or $\RSG/\RJR \simeq 0.179$. 
For the model in Fig.\ref{fig-pattern}(e), 
$\RJR$ and $J$ are estimated to be 207K and 89K, respectively. 
The ratio $J/\RJR$ is 0.43. The $g$-factor is 1.70. 
The amplitude of the spin gap is about 3.6K 
or $\RSG/\RJR \simeq 0.017$. 
In both cases, the model with given parameters $\RJR$ and $J$ 
seems to be near the critical point 
because the leading term of the finite size correction 
in the fitting function is  
$N^{-1/2}$ rather than $N^{-1}$ 
and the extrapolated values of the spin gap are very small. 
From the perturbational results, 
the origin of the spin gap for these models 
may also be described by the stripe singlet 
as well as the case in Fig.\ref{fig-pattern}(c).

Here, we discuss more quantitative aspects of the magnetic properties.
We first discuss the temperature dependence of $\chi$. 
The experimental results for the $\chi$ have not been well 
understood by the model with only the nearest-neighbor interaction, i.e.
$\RJcp = \RJcd =0$.  
In experiments, the spin gap $\RSG$, 
the peak temperature $\RTp$ and  
the temperature $T^*$ which gives a half amplitude of $\chi$ at 
$\RTp$ are 107K, 110K and 595K, respectively. 
The ratios $\RSG / \RTp$ and 
$T^*/ \RTp$ are 0.97 and 5.41, respectively. 
The peak temperature is relatively small compared to 
the spin gap and $\chi$ decays rapidly below $\RTp$. 
The numerical results for the model with the 
nearest-neighbor interactions have several different features from 
the experimental ones. 
\cite{NKatoh-JPSJ64-4105,MTroyer-PRL76-3822} 
For example, 
the ratios $\RSG / \RTp$ 
and $T^*/ \RTp$ 
for the isolated plaquette model$(\RJed =0)$ 
are 1.28 and 3.83, respectively. 
In the plaquette singlet regions, 
these ratios become small as $\RJed$ becomes large. 
While the ratios $\RSG / \RTp$ 
and $T^*/ \RTp$ 
for the isolated dimer model $(\RJep=0)$ 
are 1.60 and 3.48, respectively. 
In the dimer singlet regions, 
these ratios also become small as $\RJep$ becomes large. 
Then the ratios estimated from the experiments are no longer reproduced 
from this model. 
We consider several possible reasons of this discrepancy 
from the experimental results below $\RTp$. 
One is that the frustration due to the next-nearest-neighbor couplings 
reduces the band width of the triplet dispersion while keeps the 
strength of the spin gap, as has already been pointed out. 
\cite{KUeda-PRL76-1932}
The other is that the low-lying excitations from the 
singlet ground state are not only the triplet states 
but also the singlet states. 
For example, in the case shown in Fig.\ref{fig-pattern}(b), 
the singlet dispersion exists in the low energy region, 
as shown in Fig.\ref{fig-dispersion}(b). 
These singlet excited states are constructed from the product states 
of the lowest excited singlet state on one of the edge-shared plaquettes 
and the corner-shared dimer singlets on the others for 
the unperturbed Hamiltonian. 
Within the PE, the energy gap between the lowest singlet excitation  
and the ground state is comparable to the spin gap. 
In addition, since the singlet dispersion is nearly flat, 
the energies of the singlet excitations are lower than those of 
the triplet excitations in a wide region in the Brillouin zone. 
Then at the temperature around the singlet-singlet gap, 
$\chi$ decays rapidly since the strength of the denominator 
in eq.(\ref{eq-chi}) increases relatively  
due to the increase of the weight of the excited singlet states. 
In order to discuss the 
possibility of the existence of the low-lying excitation of the 
singlet states, 
the detailed analysis of the experimental results for the 
specific heat is needed 
because the temperature dependence of $\chi$ below the spin gap 
temperature or the neutron inelastic scattering 
do not contain the information about the singlet excitations directly.

Next, we discuss the treatment of the spin exchange couplings. 
The spin exchange couplings have been divided into two groups.  
One is strong and the others are relatively small. 
The same value has been assigned for 
the values of the other relatively small couplings. 
This weaker bonds have minor contributions to 
physical properties and do not change the essential feature 
obtained in Sec.3.  
However, within these weaker bonds, 
it is possible that the amplitude of the nearest-neighbor 
couplings becomes twice as large as that of 
the next-nearest-neighbor couplings. 
Because the number of the path for the superexchange mechanism is two 
in the nearest-neighbor case while it is one in the next-nearest-neighbor one. 
If this estimate is adopted, for example in the case shown 
in Fig.\ref{fig-pattern}(a), 
the band width of the triplet dispersion becomes narrow 
and the wavenumber giving the lowest excitation 
shifts from (0,0) to $(\pi,\pi)$ 
within the second-order PE. 
Since, in the terminology of the perturbation method, 
the transfer energy of the triplet between the nearest-neighbor 
edge-shared plaquettes is determined by 
$ \RJed -2 \RJcp $ within the first-order PE, 
it vanishes at $ \RJed =2 \RJcp $. 
The higher-order perturbation may change the wavenumber 
of the lowest excitation 
according as the sign of the transfer energy of the triplet 
or the amplitudes of the nearest and 
the next-nearest neighbor couplings.

Through this study, 
we have concentrated on the superexchange mechanism for 
the $\Rdxz$ and the $\Rdyz$ orbitals.  
In the real materials, however, 
there are several possible subtleties 
which further invalidate the assumption 
that the amplitude of the relatively small nearest-neighbor couplings 
is twice as large as that of the relatively small next-nearest ones. 
One is the contribution from the superexchange mechanism 
in the $\Rdxy$ orbital 
since the wavefunction of the occupied orbital may also have a 
$\Rdxy$ component in the real material.  
Due to this contribution, the next-nearest-neighbor couplings become  
antiferromagnetic and relevant.  
Another is the contribution from the direct exchange mechanism 
for both the $\Rdxy$ orbital and the $\Rdxz$ and $\Rdyz$ orbitals. 
Since the bonding of the 
$t_{2g}$ orbitals with 
the $\Rpz$ orbital on oxygen is not $\sigma$ bonding but 
$\pi$ bonding, 
the effect of the direct exchange mechanism 
may be relatively strong compared to that of the superexchange mechanism. 
The detailed character of the direct exchange couplings may depend on 
the symmetry of the wavefunction and the distance between the V atoms. 
The contributions from tilting of the pyramid or the Jahn-Teller distortion 
may also influence the exchange coupling. 
Detailed quantitative analysis for the spin exchange couplings 
remain for further studies.

Finally, we discuss the importance of the 
$\Rdxz$ and $\Rdyz$ orbitals in the magnetic properties for 
CaV$_4$O$_9$. 
In the real material, 
the wavefunction lying near the Fermi energy level 
may be the linear combination of the 
$d$ orbitals. 
In this context, Marini and Khomskii
\cite{SMarini-pre} 
also pointed out the importance of the orbital order. 
They discussed the effect of the crystal field 
and estimated the coefficient of 
the linear combination of the $\Rdxz$ and $\Rdyz$ orbitals 
in the wavefunction. 
The given stable orbital is tilted and has a chirality. 
They concluded that the relevant spin exchange coupling 
is $\RJed$ and the origin of the spin gap is an edge-shared dimer singlet. 
They proposed that the next strongest spin exchange couplings 
are the next-nearest-neighbor ones and 
$\RJep$ is weakly ferromagnetic. 
In this case, however, the nearest-neighbor spin correlation is 
enhanced and the out-of-phase correlation between the nearest-neighbor 
edge-shared dimer singlets should be strong 
due to the next-nearest-neighbor antiferromagnetic spin exchange couplings and 
weakly ferromagnetic $\RJep$. 
These contradict the reported experimental results.
\cite{KKodama-preprint}

Neutron measurement suggests substantial antiferromagnetic 
correlations for the next-nearest-neighbor pair of V. 
This is naturally explained by the occupation of $\Rdxy$ orbital. 
However, if only $\Rdxy$ orbitals are occupied, 
it is hard to justify the model by Kodama {\it et al.}
\cite{KKodama-preprint} 
where a much larger $\RJcp$ than $\RJcd$ is assumed. 
Aside from details it appears indeed necessary to take different 
$\RJcp$ and $\RJcd$ to reproduce the nuetron data. 
Our model of orbital order can explain why the two next-nearest-neighbor  
spin exchange couplings differ. 
However, simple model with only $\Rdxz$ and $\Rdyz$ occupations 
appears to be insufficient to explain both of the  
susceptibility and neutron results simultaneously 
as we discussed in this paper. 
All the above results imply, within this framework, that 
the wavefunction is in fact represented by a linear combination of 
the $\Rdxy$ orbital and 
the other $t_{2g}$ orbitals with the orbital order. 
We propose that the ground state 
is described as the wavefunction with large weight of 
$\Rdxy$ orbital superposed with the ordered 
$\Rdxz$ and $\Rdyz$ orbitals given in 
Fig.\ref{fig-pattern}(a). 
In this case, the basic origin of the spin gap 
may be a corner-shared plaquette singlet. 
Then the next-nearest-neighbor antiferromagnetic spin correlation 
becomes relatively large since the large weight of 
$\Rdxy$ orbital enhances $\RJcp$, 
which may explain the experimental results of 
the wavenumber dependence of the integrated 
intensity of the neutron scattering. 
Additionally, 
a strong in-phase correlation between 
the nearest-neighbor corner-shared plaquette singlets increases 
because of the existence of $\RJed$ due to the presence of the orbital order.  
Then the wavenumber of the lowest triplets may become $(0,0)$, 
which is consistent with the recent experimental results. 
\cite{KKodama-preprint} 
If this situation is true,  
small plaquette which is constructed by four oxygens 
without a V atom at the center may rotate  
or some other specific type of the lattice distortion may occur  
in order to favor the given orbital order.
To obtain the information on the weight of these orbitals, 
a detailed analysis of the lattice structure
is needed, especially at very low temperatures. 
Quantitative and detailed theoretical analyses 
of this proposal remain for further studies.

The orbital order effect may also be observed in the 
temperature dependence of $\chi$ for CaV$_2$O$_5$. 
\cite{HIwase-JPSJ65-2397}
This material also shows the spin gap behavior. 
The spin gap and the peak temperature are 
about 464K and 400K, respectively. 
One possible mechanism of the spin gap formation 
may be due to the ladder structure. 
Since it is assumed that $d$ electron is localized at the 
$\Rdxy$ orbital, 
the spin exchange couplings $J$ in the leg and 
$J_{\perp}$ in the rung have almost the same values. 
Numerical analysis of $\chi$ for the AFH model on the 
ladder lattice has led to the result that 
the ratio $\RSG / \RTp$ is almost 0.5 at $J=J_{\perp}$.
\cite{MTroyer-PRB50-13515}
However, the ratio obtained in the experiments is 1.16, 
which is different from the theoretical prediction. 
If the localized orbital on V atom is represented by the 
linear combination of the $\Rdxy$ and 
one of the other $t_{2g}$ orbitals 
which extends to the oxygen on the rung due to the orbital order, 
$J_{\perp}$ becomes larger than $J$. 
Since the ratio $\RSG / \RTp$ is 1.60 in the dimer limit $(J=0)$, 
it may become 1.16 as $J$ increases from zero. 
Quantum or thermal fluctuation of the orbital degree of freedom 
also changes a quantitative behavior of $\chi$ at finite temperature. 
We propose that the orbital order of $\Rdxz$ and $\Rdyz$ orbitals 
with partially fillings of $t_{2g}$ orbitals is also a promising 
explanation for rather puzzling experimental results in CaV$_2$O$_5$.  
The quantitative study for this problem remains for further study.

\section{Summary}

We have investigated orbital order effects in 
magnetic properties of CaV$_4$O$_9$. 
Several possible models with the orbital order have been considered. 
The amplitudes of the spin exchange couplings for each model 
are determined in order to reproduce the temperature dependence of $\chi$ 
in experiments. 
The origins of the spin gap are not necessarily the 
same as the originally proposed plaquette singlet 
but are ascribed to the generalized four-site singlet.  
By using the estimated values of the spin exchange couplings, 
the dispersion of the triplet excitations and 
the spin-spin correlation corresponding to the integrated scattering intensity 
have been calculated within the PE. 
The strong in-phase correlation between the nearest-neighbor 
pair of four-site singlets and 
the large antiferromagnetic spin correlation between 
the next nearest-neighbor spins are 
minimal requirements for the explanations of 
the experimental results in the neutron scattering.
Although 
the wavenumber dependences of the triplet dispersion 
and the spin-spin correlation for our models 
do not show completely satisfactory agreement with those 
obtained by the presently available experimental data, 
they are much improved from the single-orbital cases.  
We further propose the mechanism 
required to explain each feature of the experimental results. 
We suggest that the order of the $\Rdxz$ and $\Rdyz$ orbitals 
hybridized with uniform and partial occupation of $\Rdxy$ orbital  
may explain the experimental results. 
We have also discussed that effects of the orbital order may be observed in 
other vanadium oxide compounds such as CaV$_2$O$_5$. 
These strongly suggest that 
the role of the orbital order is important 
in understanding the magnetic properties of the vanadium oxide compounds 
with the spin gap.

\section*{Acknowledgements}
We would like to thank M. Sato, S.Taniguchi, K. Kodama
for useful discussions and comments. 
We have used a part of the codes provided by H. Nishimori in TITPACK Ver.2. 
A part of the computation has been performed on VPP500 at 
the Supercomputer Center of the Institute for Solid State Physics, 
Univ.of Tokyo. 
This work is financially supported by a Grant-in-Aid for 
Scientific Research in Priority Areas 
``Anomalous Metallic States near the Mott Transition''.  

\appendix
\section{Triplet Dispersion and Wavefunction by Perturbation Expansion}

We introduce the procedure for calculating 
the energy dispersion of the lowest triplet band and 
the wavefunctions of the singlet ground state and 
the triplet band. 

We first consider the model shown in Fig.\ref{fig-pattern}(a).
The unperturbed Hamiltonian has the terms with $\RJep$ 
on the edge-shared plaquette, 
while the other terms are treated as the perturbed ones. 
The ground state of the edge-shared plaquette with 
only $\RJep$ is the RVB state $|\phi_{\mbox{g}}\rangle$. 
The energy of this state is $-2 \RJep$. 
Then the ground state of the unperturbed Hamiltonian is 
the product state of $|\phi_{\mbox{g}}\rangle$ written as
%%%
%%%
\begin{equation}
|\Phi_{\mbox{g}}\rangle_{(a)}=\prod_{i}|\phi_{\mbox{g}}
(\mbox{\boldmath{$r$}}_i)\rangle.
\end{equation}
%%%
%%%
The lowest triplet states of the edge-shared plaquette 
$|\psi\rangle$ with $S_z=-1,0,1$ are the hopping state of the 
triplet pair in the RVB state. The energy of these states is $-\RJep$. 
Then the wavefunctions of the lowest triplet band are constructed by 
$|\psi\rangle$ on one of the edge-shared plaquette 
and $|\phi_{\mbox{g}}\rangle$ on the others. 
The degeneracy is lifted by the first-order perturbation. 
The wavefunctions of these triplet states are written as
%%%
%%%
\begin{equation}
|\Psi(\Rbk)\rangle_{(a)}=
\frac{1}{\sqrt{N_{\mbox{\small p}}}}\sum_i |\psi
(\mbox{\boldmath{$r$}}_i)\rangle
\prod_{j\ne i} |\phi_{\mbox{g}}
(\mbox{\boldmath{$r$}}_j)\rangle
\mbox{e}^{\mbox{i}
\mbox{\boldmath{$k \cdot r$}}_i
}, 
\end{equation}
%%%
%%%
where $N_p$ is the number of the edge-shared plaquette.
Within the second-order PE, 
the energy difference between the triplet states 
and the singlet ground state is represented as
%%%
%%%
\begin{eqnarray*}
\lefteqn{
\mbox{\hspace{-2cm}}
\RSG (\Rbk)_{(a)}=
\RJep -\frac{115}{864}\frac{J^2}{\RJep}
-(\frac{J}{3}+\frac{J^2}{72J_{\RJep}})
(\cos k_x +\cos k_y)
}
\nonumber\\
& &+\frac{J^2}{108 \RJep}
(\cos 2 k_x + \cos 2 k_y)
-\frac{J^2}{9 \RJep} \cos k_x  \cos k_y. 
\end{eqnarray*}
%%%
%%%

Next, we consider the model shown in Fig.\ref{fig-pattern}(b). 
The unperturbed Hamiltonian has the terms with 
$\RJep$ and $\RJcd$ on the edge-shared plaquette 
while the other terms are the perturbed ones. 
Since we take $\RJcd  > \RJep $, for example, 
$\RJep =0.9 \RJcd$, 
the ground state of the edge-shared plaquette is not 
the $|\phi_{\mbox{g}}\rangle$ but the product state  
$|\xi_{\mbox{g}}\rangle$ 
of the dimer pair on each $\RJcd$ bond. 
The energy of this state is $-1.5 \RJcd$. 
Then the ground state of the unperturbed Hamiltonian is the product state of 
$|\xi_{\mbox{g}}\rangle$ written as 
%%%
%%%
\begin{equation}
|\Phi_{\mbox{g}}\rangle_{(b)}=\prod_i |\xi_{\mbox{g}}
(\mbox{\boldmath{$r$}}_i)\rangle. 
\end{equation}
%%%
%%%
The wavefunctions of the lowest triplet states on the edge-shared plaquette 
are constructed by two states 
$|\eta\rangle$ and $|\zeta\rangle$ 
with $S_z=-1,0,1$ 
which are the product states of 
the triplet pair on one of the $\RJcd$ bond 
and the dimer pair on the others. 
Then the wavefunction of the lowest triplet bands are 
constructed from one of the triplets 
$|\eta\rangle$ or $|\zeta\rangle$ 
on one of the edge-shared plaquette and 
$|\xi_{\mbox{g}}\rangle$ on the others. 
The first-order perturbation lifts the degeneracy. 
The bonding and anti-bonding states of the triplets appear, 
which are written as 
%%%
%%%
\begin{equation}
|\Psi_{\pm}(\Rbk)\rangle_{(b)}=
\frac{1}{\sqrt{N_{\mbox{\small p}}}}\sum_{\Rbk}
(
\frac{1}{\sqrt{10\mp 4\sqrt{5}}}
|\eta(\mbox{\boldmath{$r$}}_i)\rangle
+
\frac{-2\pm \sqrt{5}}{\sqrt{10\mp 4\sqrt{5}}}
|\zeta(\mbox{\boldmath{$r$}}_i)\rangle
)
\prod_{j\ne i}
|\xi_{\mbox{g}}(\mbox{\boldmath{$r$}}_j)\rangle
\mbox{e}^{\mbox{i}
\mbox{\boldmath{$k \cdot r$}}_i}, 
\end{equation}
%%%
%%%
in the parameter regions $0 \le k_x \le k_y \le \pi$.  
Here $\RJcp$ and $\RJed$ are both taken as $J$.
%%%
%%%
%%%\begin{equation}
%%%A_{\pm}=\frac{1}{\sqrt{2}}\frac{2\gamma}
%%%{\sqrt{(\alpha-\beta)^2+4\gamma^2\pm(-\alpha+\beta)
%%%\sqrt{(\alpha-\beta)^2+4\gamma^2}
%%%}},
%%%\end{equation}
%%%\begin{equation}
%%%B_{\pm}=\frac{1}{\sqrt{2}}
%%%\frac
%%%{
%%%-\alpha+\beta\pm\sqrt{(\alpha-\beta)^2+4\gamma^2}
%%%}
%%%{
%%%\sqrt{(\alpha-\beta)^2+4\gamma^2\pm(-\alpha+\beta)
%%%\sqrt{(\alpha-\beta)^2+4\gamma^2}
%%%}
%%%},
%%%\end{equation}
%%%
%%%
%%%where
%%%
%%%
%%%\begin{eqnarray}
%%%\alpha&=&-\frac{3}{4}J\cos k_x + \frac{1}{4}J\cos k_y,
%%%\nonumber\\
%%%\beta&=&\frac{1}{4}J\cos k_x - \frac{3}{4}J\cos k_y,
%%%\nonumber\\
%%%\gamma&=&-\frac{1}{4}J\cos k_x + \frac{1}{4}J\cos k_y,
%%%\end{eqnarray}
%%%
%%%
%%%respectively. 
The energy difference between these triplet states and 
the ground state is obtained as 
%%%
%%%
%%%\begin{eqnarray*}
%%%\lefteqn{
%%%\Delta_s(\mbox{\boldmath{$k$}})
%%%_{\pm}=
%%%J_{\mbox{cd}}
%%%+\frac{3J^2}{8J_{\mbox{cd}}}
%%%-\frac{J^2}{16( J_{\mbox{cd}}-J_{\mbox{ep}})}
%%%-\frac{3J^2}{8(2J_{\mbox{cd}}-J_{\mbox{ep}})}
%%%-\frac{5J^2}{8(2J_{\mbox{cd}}+J_{\mbox{ep}})}
%%%}
%%%\nonumber\\
%%%& &
%%%\left\{
%%%-\frac{-1\mp \sqrt{5}}{4}J 
%%%-\frac{1}{J_{\mbox{cd}}}
%%%\left[
%%%\frac{J^2}{4}
%%%(A_{\pm}^2-B_{\pm}^2)
%%%+
%%%\frac{J^2}{8}
%%%(A_{\pm}+B_{\pm})^2
%%%\right]
%%%\right\}
%%%\cos k_x
%%%\nonumber\\
%%%& &
%%%\left\{
%%%-\frac{-1\pm \sqrt{5}}{4}J 
%%%-\frac{1}{J_{\mbox{cd}}}
%%%\left[
%%%\frac{J^2}{4}
%%%(-A_{\pm}^2+B_{\pm}^2)
%%%+\frac{J^2}{8}
%%%(A_{\pm}-B_{\pm})^2
%%%\right]
%%%\right\}
%%%\cos k_y,
%%%\end{eqnarray*}
%%%
%%%
\begin{eqnarray*}
\lefteqn{\mbox{\hspace{-2cm}}
\RSG(\Rbk)_{\pm}= \RJcd
+\frac{3J^2}{8 \RJcd}
-\frac{J^2}{16( \RJcd - \RJep)}
-\frac{3J^2}{8(2 \RJcd - \RJep)}
-\frac{5J^2}{8(2 \RJcd + \RJep)}
}
\nonumber\\
& &+
\left(
\frac{J}{4}+\frac{J^2}{8 \RJcd}
\right)
\left[
(-1\mp\sqrt{5})\cos k_x
+
(-1\pm\sqrt{5})\cos k_y
\right].
\end{eqnarray*}
%%%
%%%
In this model, the lowest singlet band is also calculated. 
In a similar way, the energy difference between the 
lowest singlet band and the singlet ground state is obtained as 
%%%
%%%
%%%\begin{eqnarray*}
%%%\lefteqn{
\begin{equation}
\RSG'(\Rbk)=
2( \RJcd - \RJep )
+\frac{9J^2}{8 \RJcd}
-\frac{3J^2}{8 \RJep}
-\frac{J^2}{2( \RJcd  + \RJep)}
%%%
%%%\nonumber\\
%%%& &
+\frac{J^2}{8 \RJep}
(\cos k_x+\cos k_y).
\end{equation}
%%%\end{eqnarray*}
%%%
%%%

Next we introduce the wavefunctions for the models 
shown in Fig.\ref{fig-pattern}(c).
In the case shown in Fig.\ref{fig-pattern}(c), 
the unperturbed Hamiltonians are taken as the part with 
the relevant next-nearest-neighbor spin exchange couplings, 
while the other part with the relevant nearest-neighbor ones are the 
perturbed ones 
because we suggest that another possible origin of the spin gap 
is described as the stripe singlet in this case. 
Since the magnetic unit cell has two stripes, 
the ground state of the unperturbed Hamiltonians 
$|\Phi_{\mbox{g}}\rangle_{(c)}$ is represented by 
the product state of the stripe singlet 
$|\phi'_{\mbox{g}}(\mbox{\boldmath{$r$}}_i)\rangle_{1,2}$
on each stripe written as
%%%
%%%
\begin{equation}
|\Phi_{\mbox{g}}\rangle_{(c)}=\prod_i 
|\phi'_{\mbox{g}}(\mbox{\boldmath{$r$}}_i)\rangle_1
|\phi'_{\mbox{g}}(\mbox{\boldmath{$r$}}_i)\rangle_2,
\end{equation}
%%%
%%%
where the indices 1 and 2 represent 
the one stripe and the other stripe in the magnetic unit cell, respectively. 
The lowest triplet states of the magnetic unit cell for the unperturbed 
Hamiltonian are six-fold degenerated i.e.
$|\psi'\rangle_1 |\phi'_{\mbox{g}}\rangle_2$ and 
$|\phi'_{\mbox{g}}\rangle_1 |\psi'\rangle_2$ 
with $S_z=-1,0,1$. 
Then the bonding and anti-bonding triplet bands appear 
by the PE. 
Within the first-order perturbation, 
the wavefunctions are obtained as
%%%
%%%
\begin{equation}
|\Psi_{\pm}(\Rbk)\rangle_{(c)}=
\frac{1}{\sqrt{2N_{\mbox{\small s}}}}
\sum_i
(|\psi'({\mbox{\boldmath{$r$}}_i})\rangle_1 
 |\phi'_{\mbox{g}}({\mbox{\boldmath{$r$}}_i})\rangle_2
\pm
 |\phi'_{\mbox{g}}({\mbox{\boldmath{$r$}}_i})\rangle_1
 |\psi'({\mbox{\boldmath{$r$}}_i})\rangle_2
)
\prod_{j\ne i}
|\phi'_{\mbox{g}}(\mbox{\boldmath{$r$}}_j)\rangle_1
|\phi'_{\mbox{g}}(\mbox{\boldmath{$r$}}_j)\rangle_2,
\mbox{e}^{\mbox{i}
\mbox{\boldmath{$k \cdot r$}}_i}, 
\end{equation} 
%%%
%%%
where $N_{\mbox{\small s}}$ is the number of the stripe.
We calculate the energy difference between these triplet bands 
and the ground state within the second-order PE  
written as 
%%%
%%%
\begin{equation}
\RSG(\Rbk)_{\pm}=J'[A_1\pm A_2
(-\cos k_x + \cos k_y)], 
\end{equation}
where the parameters $A_1$ and $A_2$ are 0.601 and 0.031, respectively 
at $J/J'=0.38$. 
Here the component $k_x$ and $k_y$ are represented by the 
vectors 
$\tilde{\mbox{\boldmath $a$}_1}$ 
and 
$\tilde{\mbox{\boldmath $b$}_1}$, 
respectively.

\appendix
\section{Ground State Energy for Model in Fig.
\ref{fig-pattern}(c)}

We calculate the ground state energy per site 
by the second-order PE 
to investigate the possible mechanism of the spin gap formation 
for the model shown in Fig.\ref{fig-pattern}(c). 
The first case is that the unperturbed Hamiltonian 
is the term with $\RJep$ and the other terms are 
the perturbed Hamiltonian. 
In this case, the origin of the spin gap for the 
unperturbed Hamiltonian is the edge-shared plaquette singlet. 
The strength of the other spin exchange couplings is taken as $J$. 
The ground state energy per site is obtained as
%%%
%%%
\begin{equation}
\epsilon_{\mbox{g}}=-\frac{1}{2} \RJep +\frac{1}{16}J
-\frac{11}{576}\frac{J^2}{\RJep}.
\end{equation}
%%%
%%%
The second case is that the 
unperturbed Hamiltonian is the term with $\RJed$ 
while the others are the perturbed ones.
The strength of the other spin exchange couplings is taken as $J$. 
In this case, the origin of the spin gap is the dimer singlet. 
The ground state energy per site is obtained as
%%%
%%%
\begin{equation}
\epsilon_{\mbox{g}}=-\frac{3}{8} \RJed -\frac{9}{128}
\frac{J^2}{\RJed}.
\end{equation}
%%%
%%%
The third case is that the term with the next-nearest-neighbor
spin exchange coupling $J'$ is the unperturbed Hamiltonian 
while the others are the perturbed ones. 
Similarly, 
the strength of the other spin exchange coupling is taken as $J$. 
In this case, the origin of the spin gap is the stripe singlet.
The ground state energy per site is obtained as
%%%
%%%
\begin{equation}
\epsilon_{\mbox{g}}=-\frac{3+2\sqrt{3}}{16}J'
-\frac{J^2}{J'}\left(\frac{A}{8}+\frac{B}{4}\right),
\end{equation}
%%%
%%%
where the parameters $A$ and $B$ are written as
%%%
%%%
\begin{equation}
A=\frac{1}{3(1+\sqrt{3})},
\end{equation}
\begin{eqnarray*}
\lefteqn{
B=\frac{1}{768}
\left[
\frac{(4-\sqrt{2}-2\sqrt{3}+\sqrt{6})^2}{1-\sqrt{2}+\sqrt{3}}
+
\frac{(2-3\sqrt{2}+\sqrt{6})^2}{1+\sqrt{3}}
+
\frac{8(\sqrt{3}-1)^2(2-\sqrt{2})}{2-\sqrt{2}+2\sqrt{3}}
\right.
}
\nonumber\\
& &
\left.
+
\frac{(-4-\sqrt{2}+2\sqrt{3}+\sqrt{6})^2}{1+\sqrt{2}+\sqrt{3}}
+
\frac{(2+3\sqrt{2}+\sqrt{6})^2}{1+\sqrt{3}}
+
\frac{8(\sqrt{3}-1)^2(2+\sqrt{2})}{2+\sqrt{2}+2\sqrt{3}}
\right],
\end{eqnarray*}
respectively. 
When the parameters $\RJep$ and $\RJed$ are taken as $J$, 
the $J/(J+J')$ dependence of the ground state energy per site 
is obtained, as shown in Fig.\ref{fig-gse}.

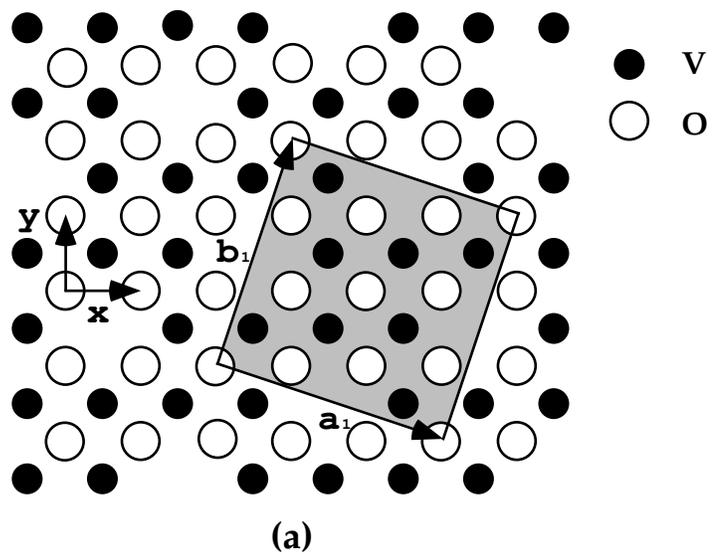
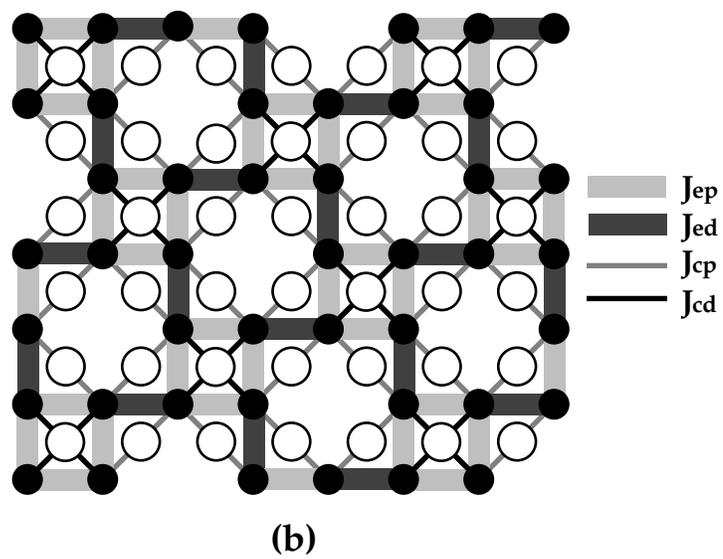
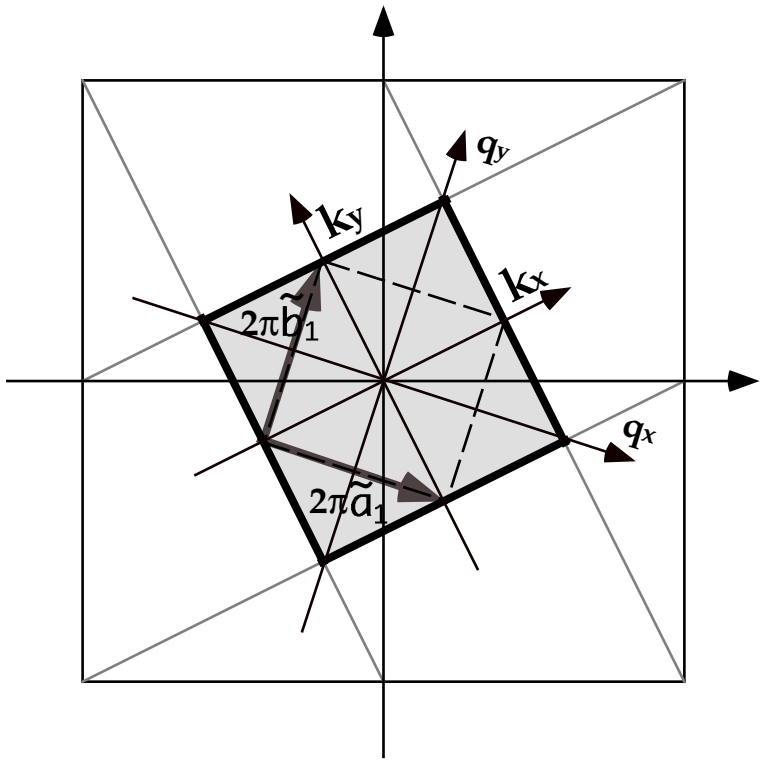
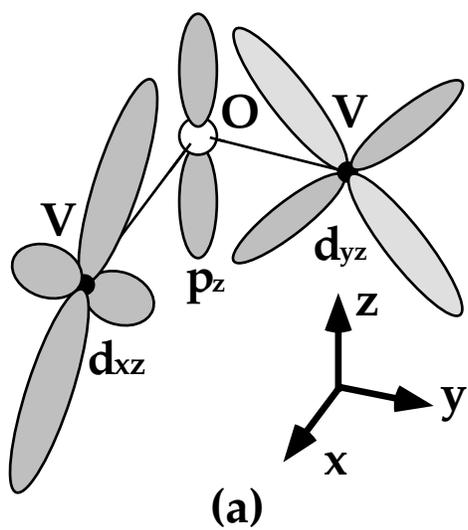
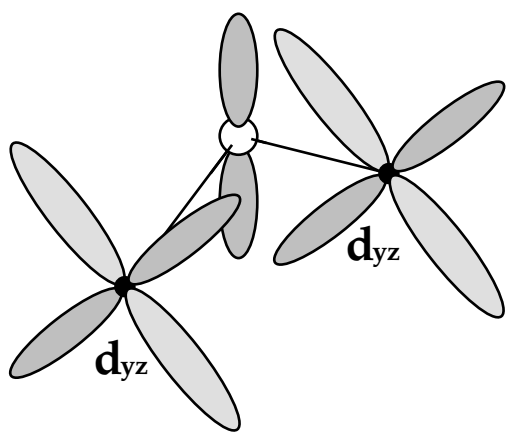
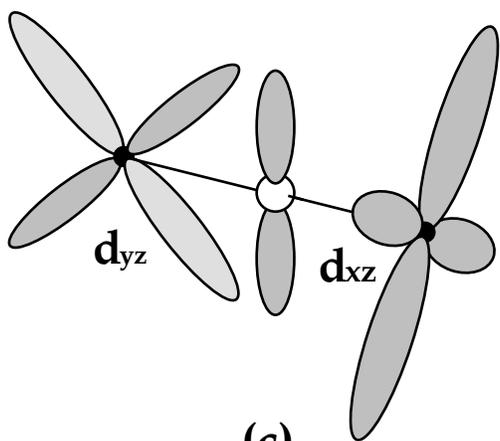
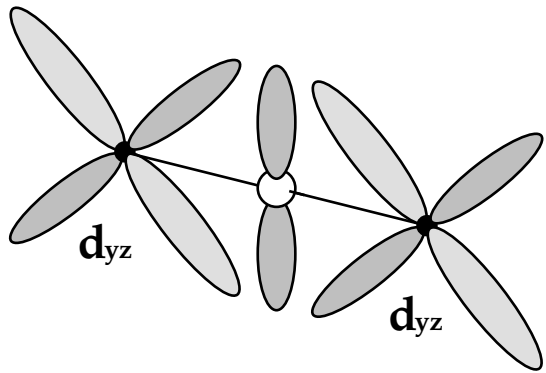
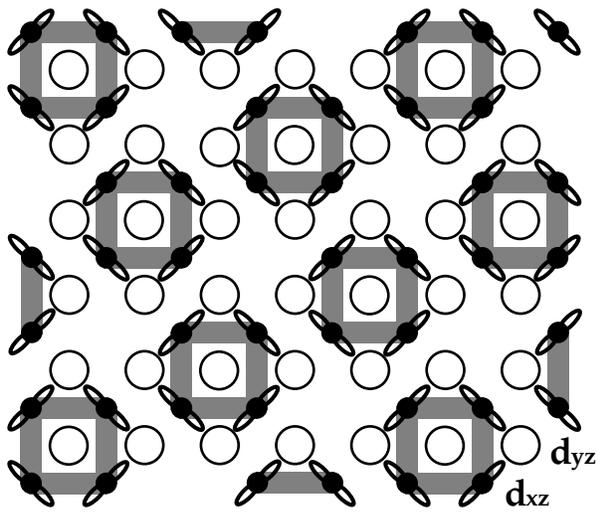
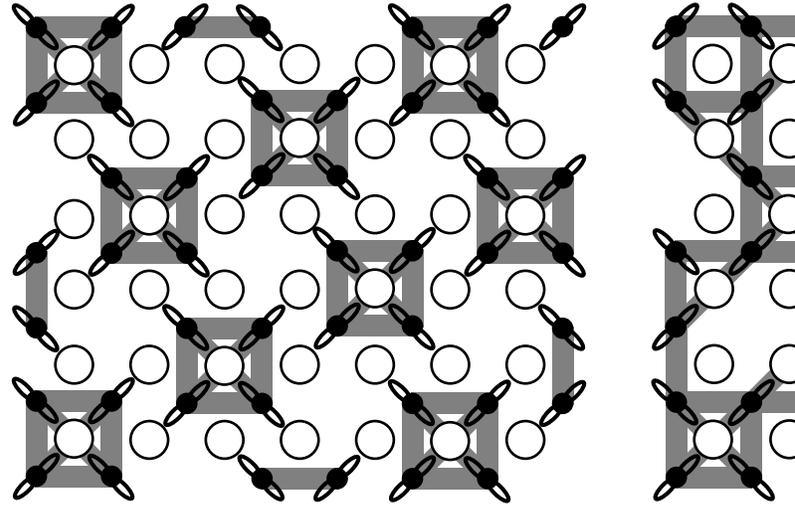
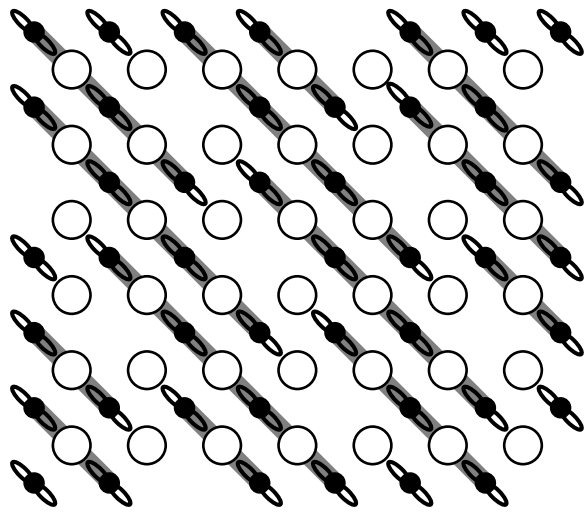
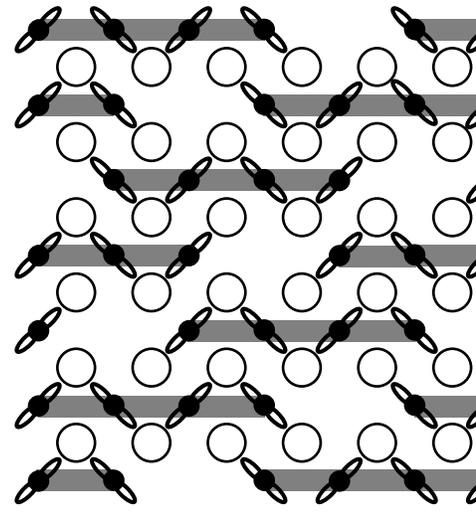
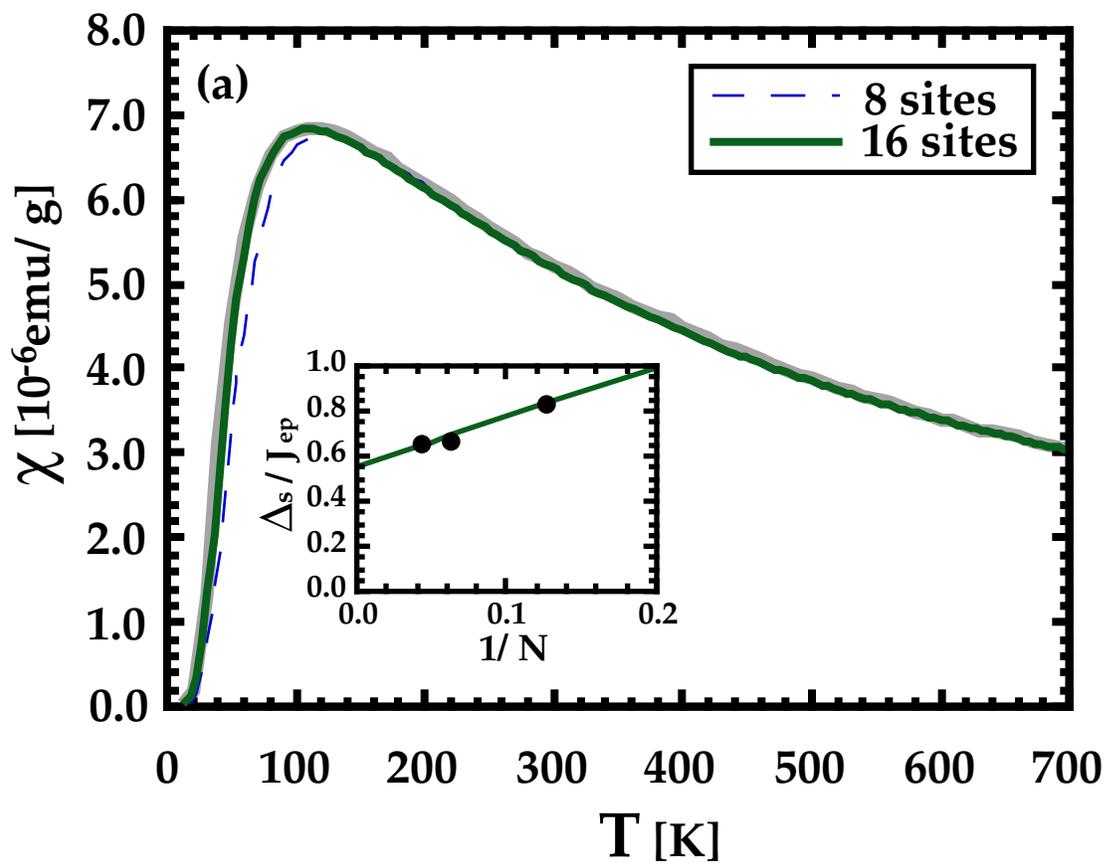
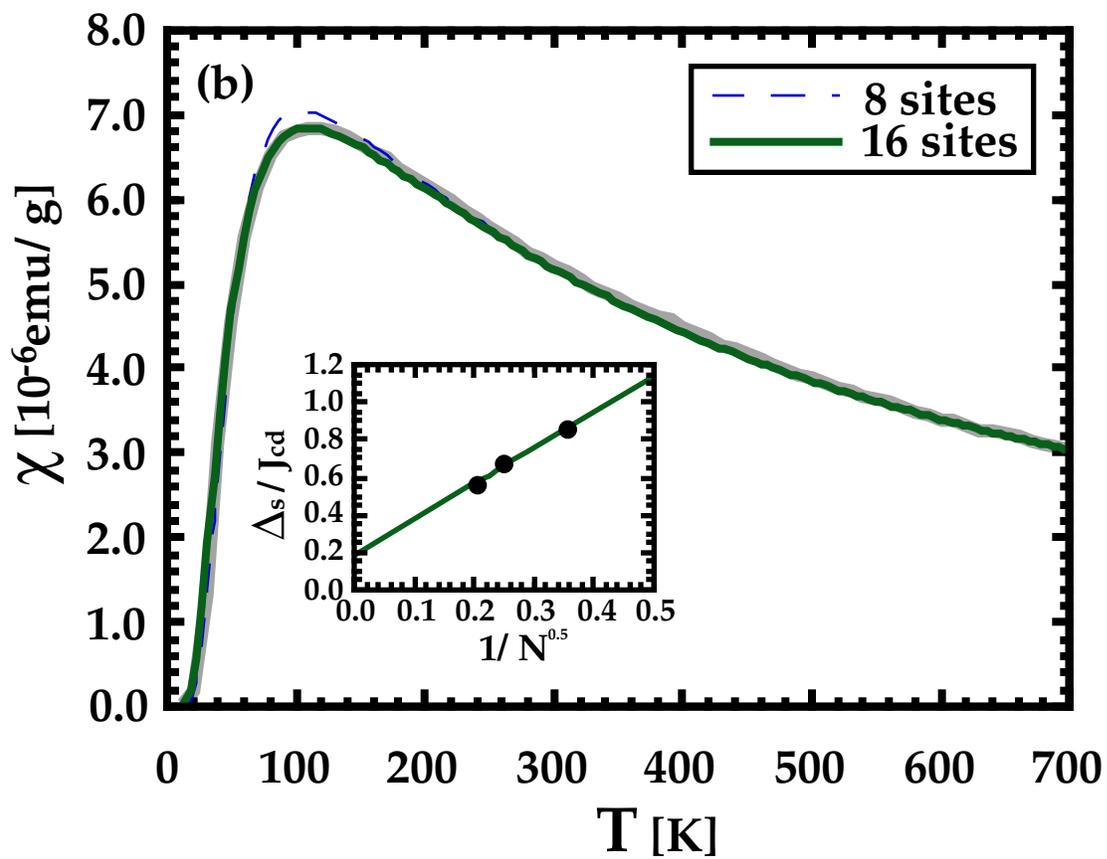
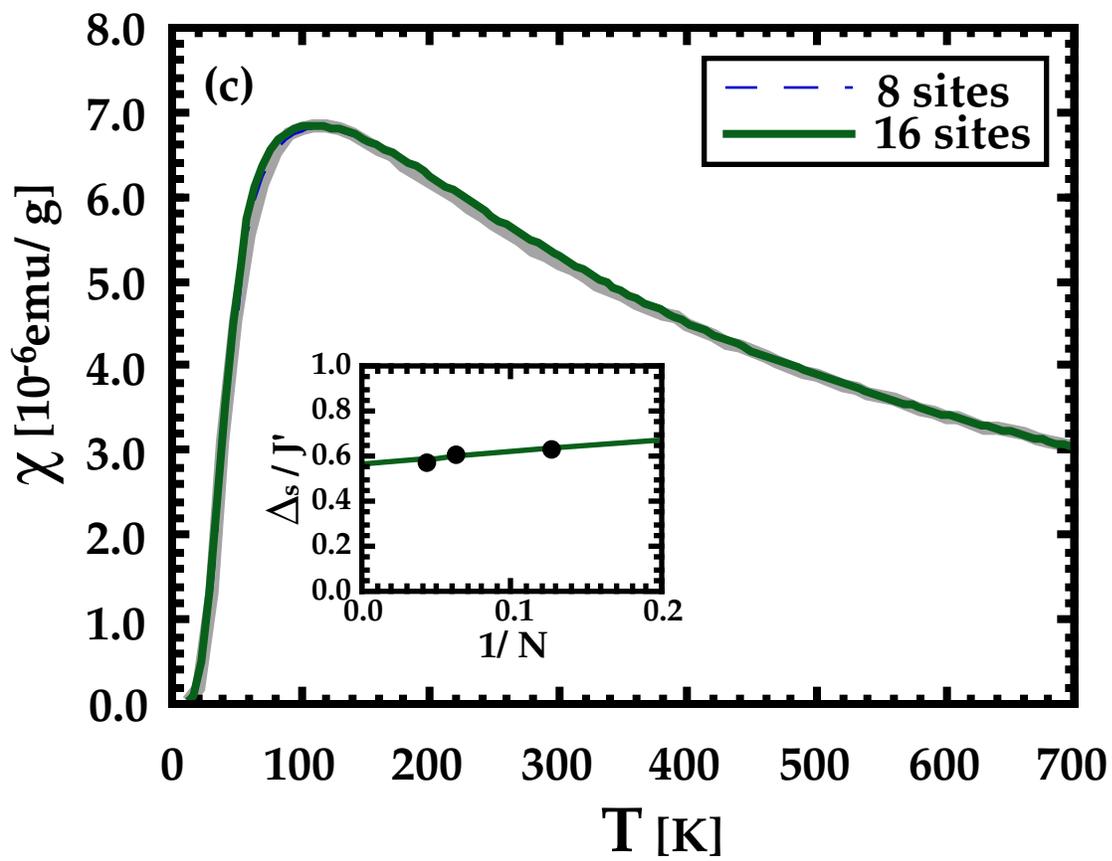
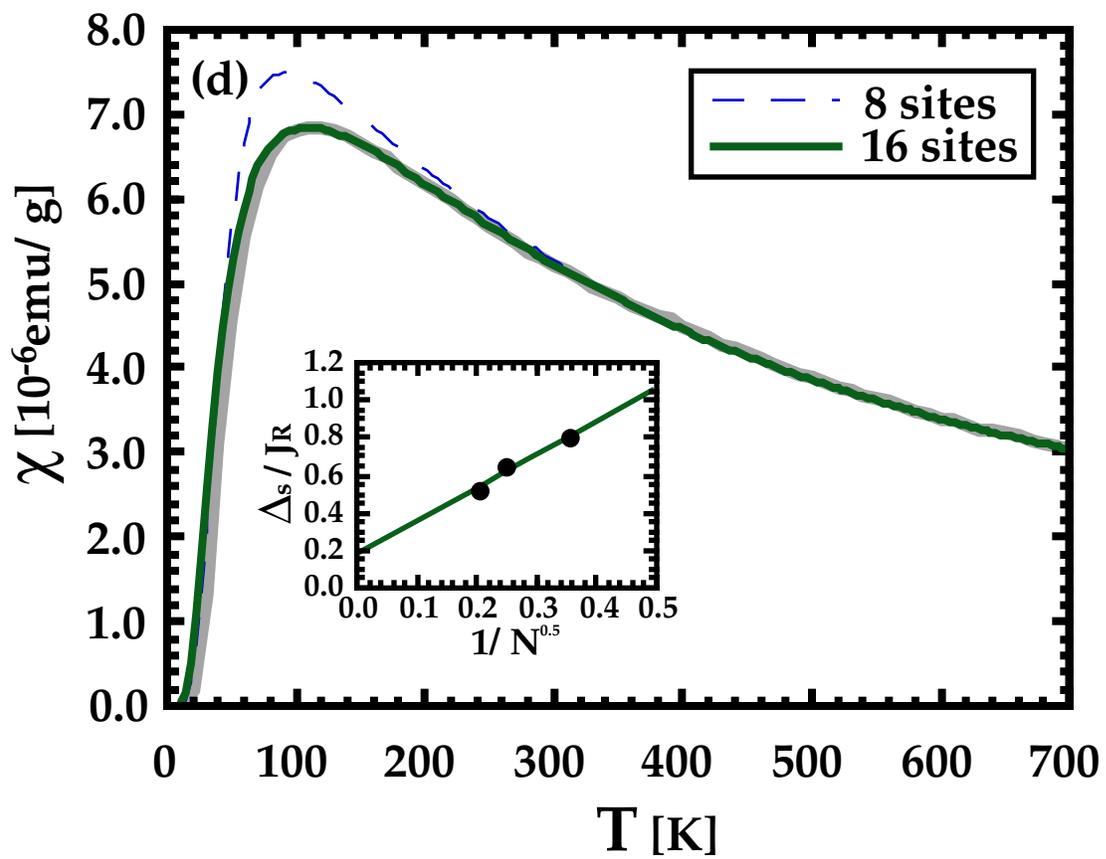
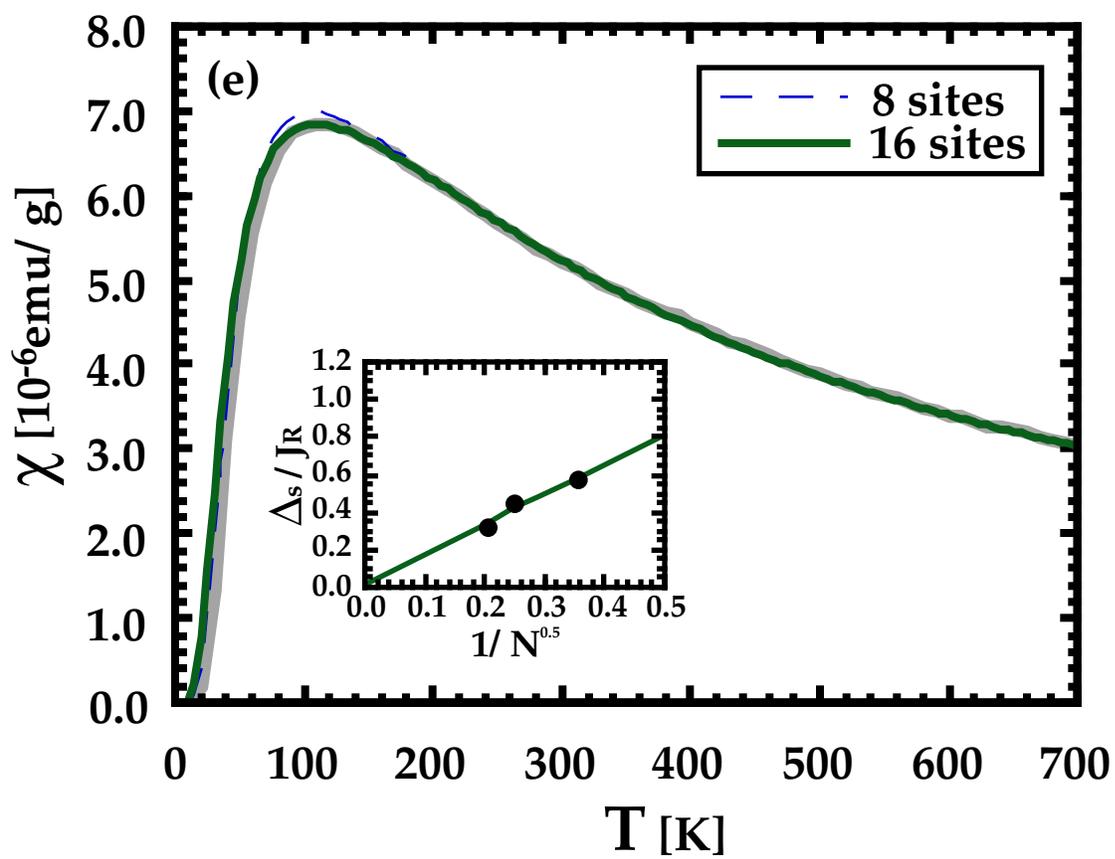
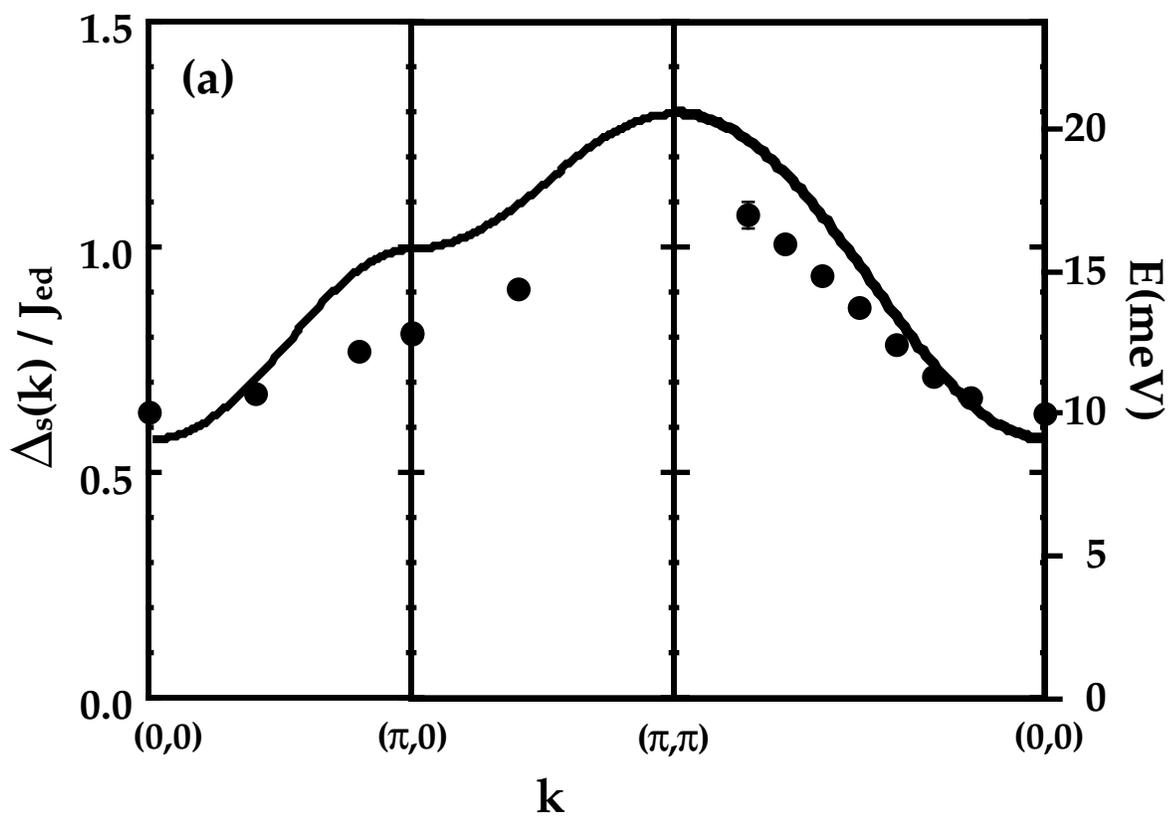
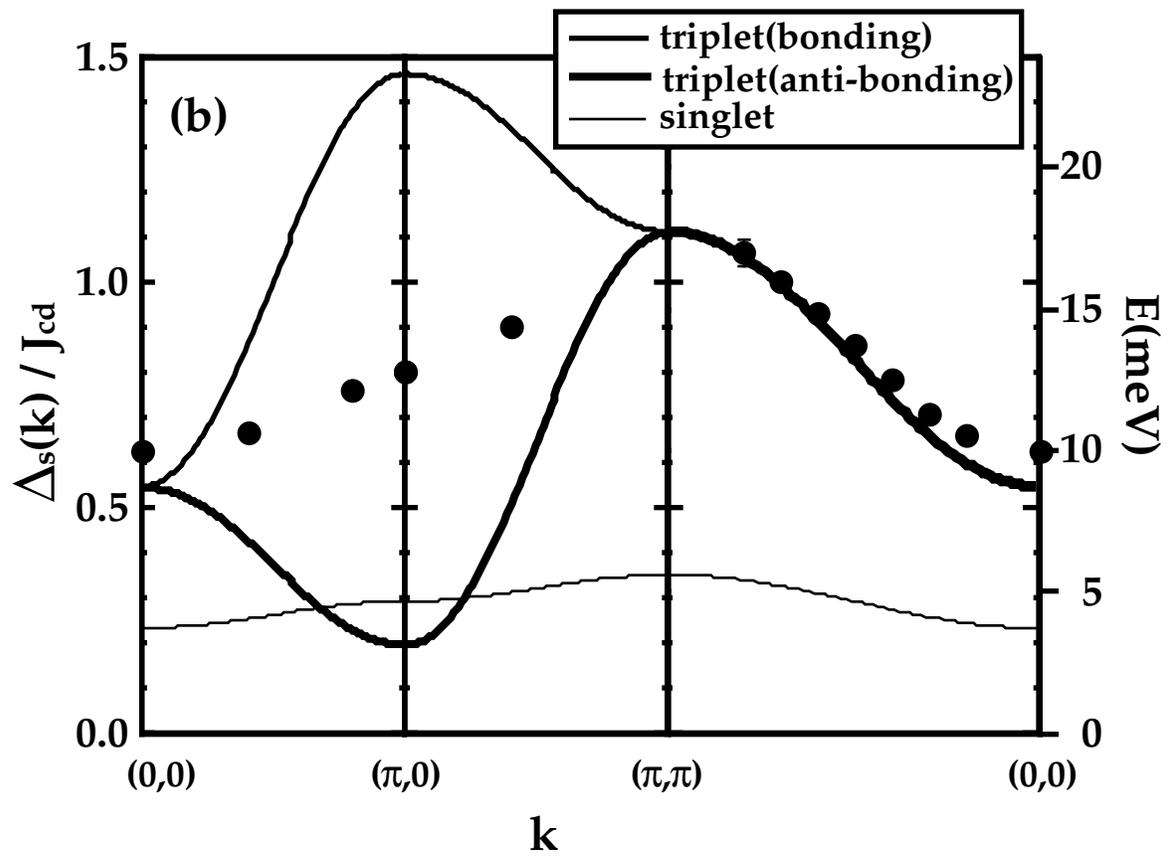
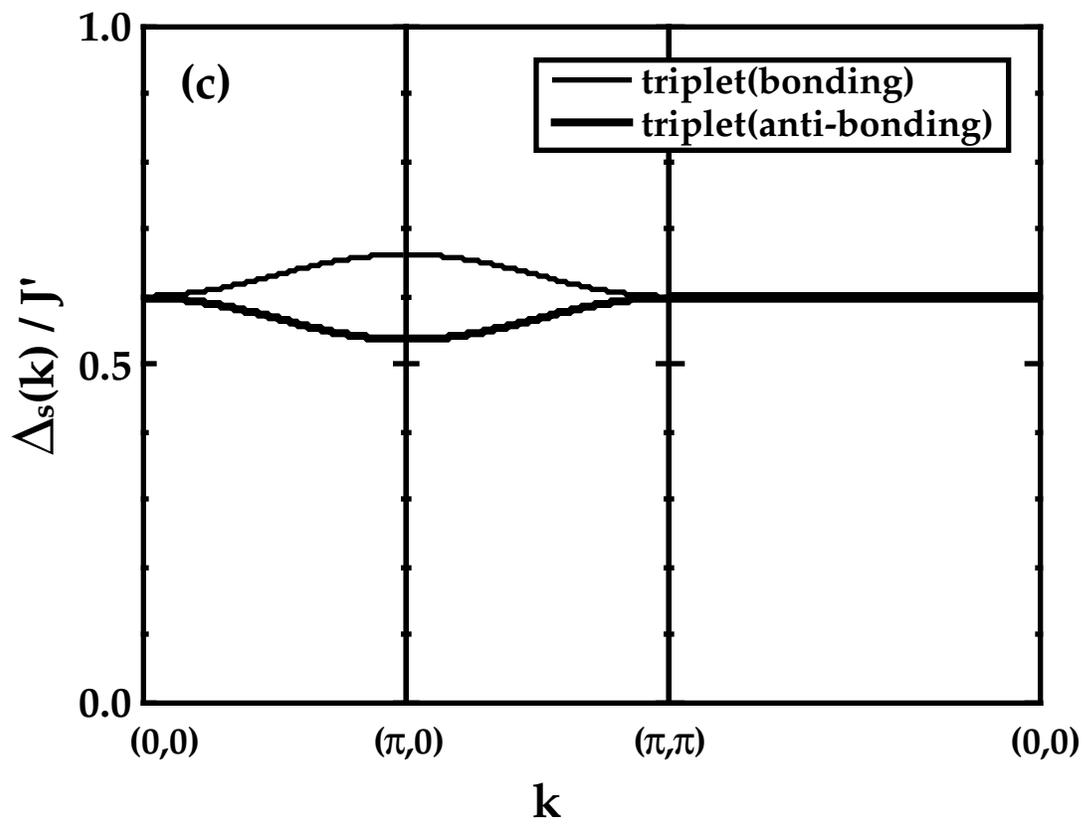
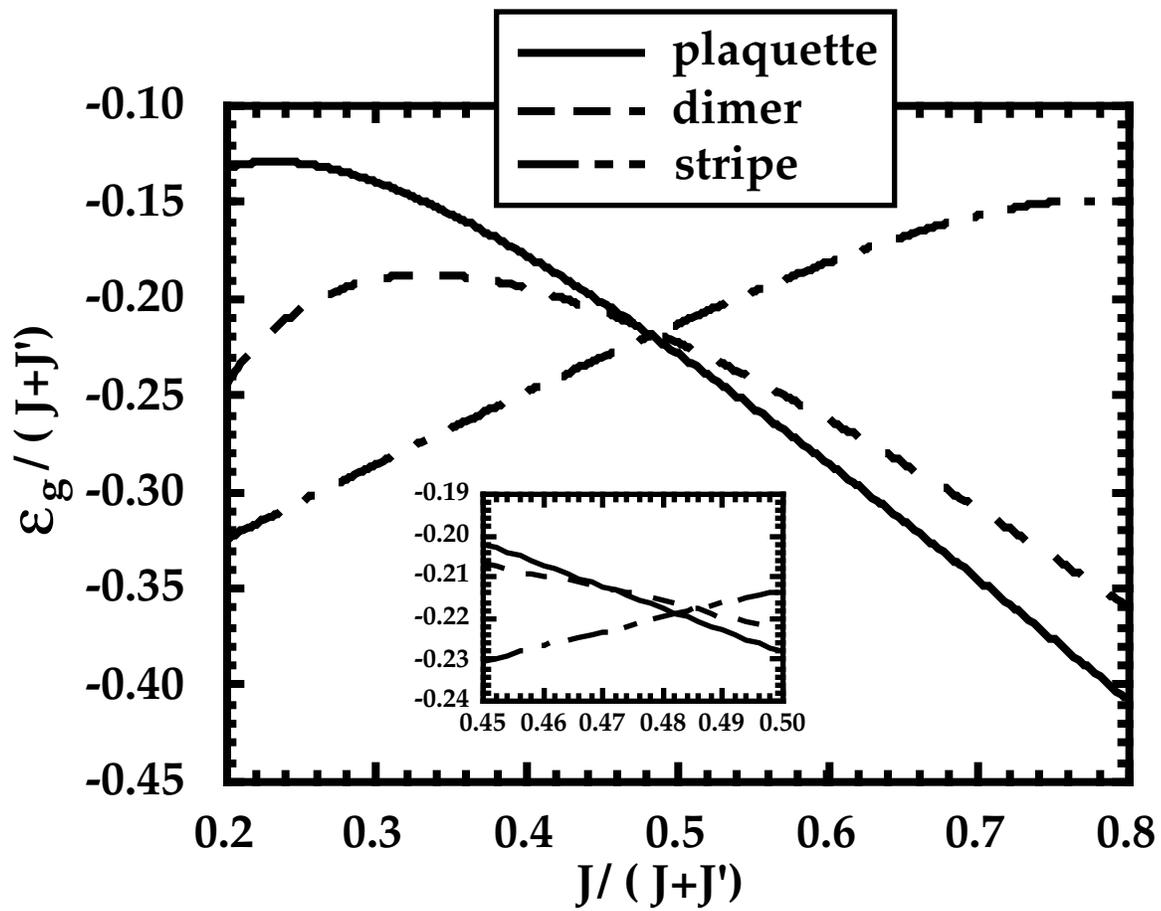
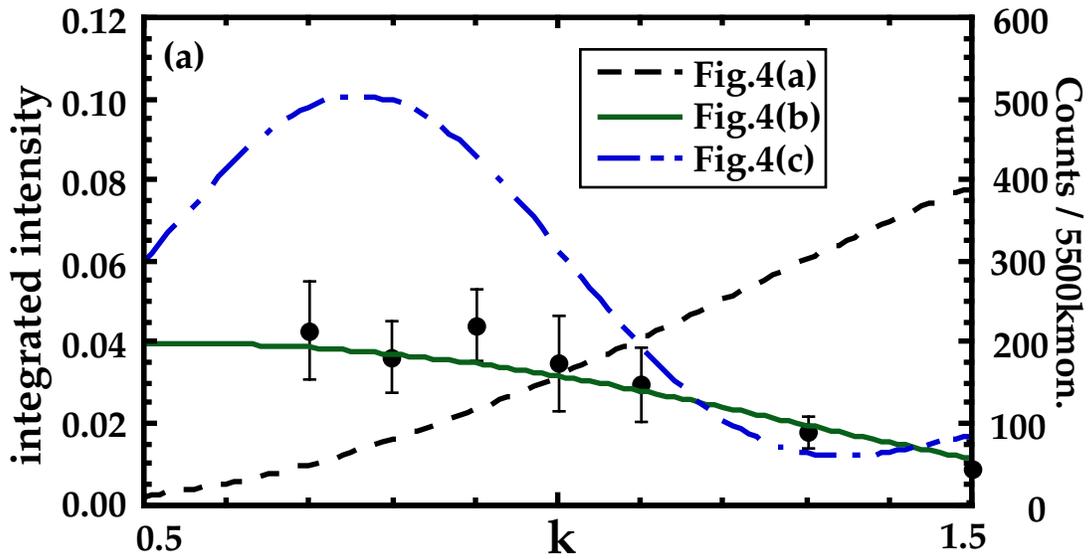
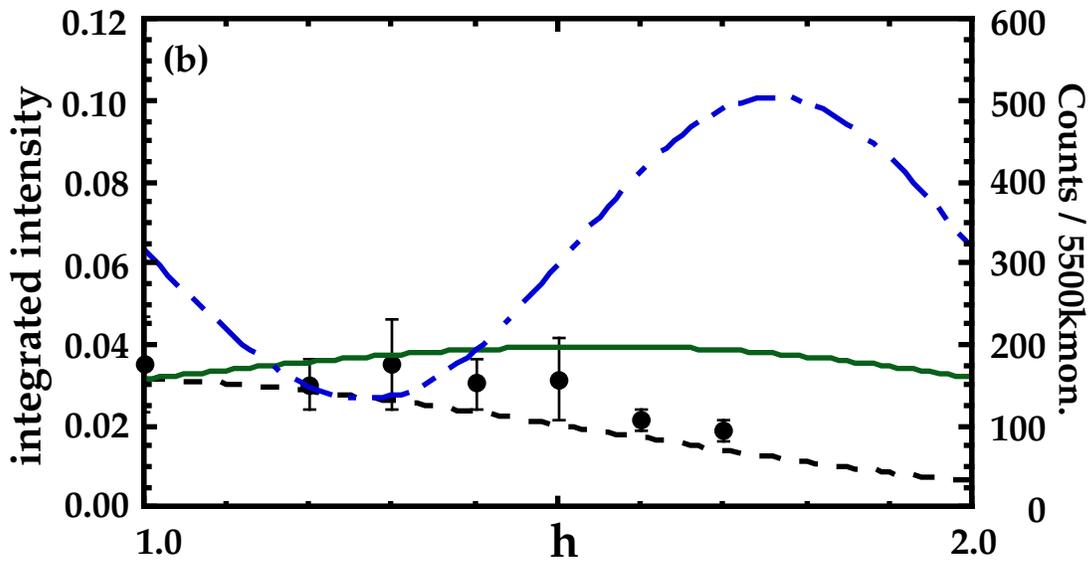
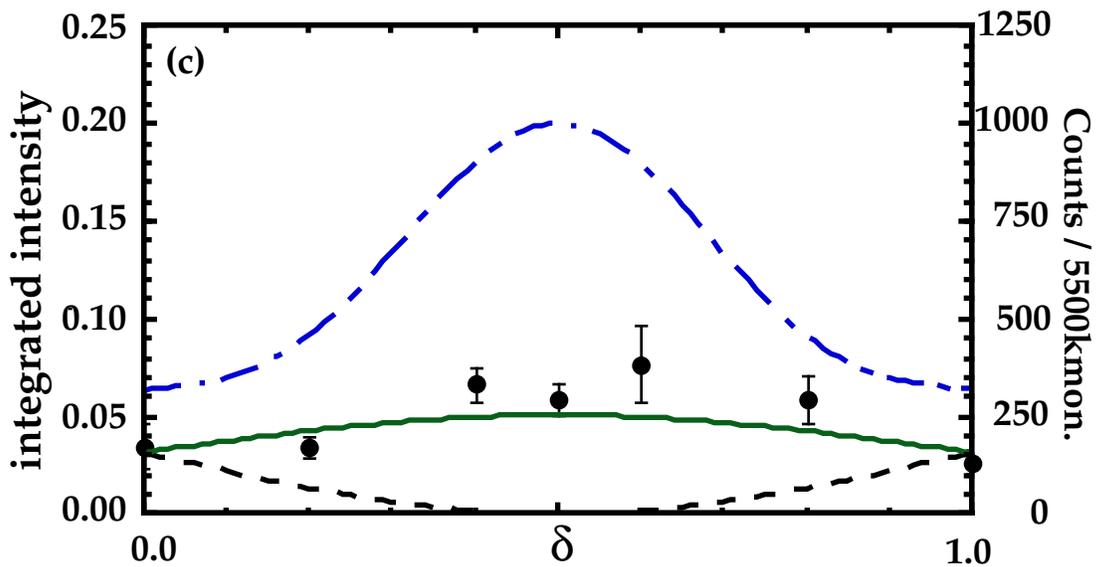
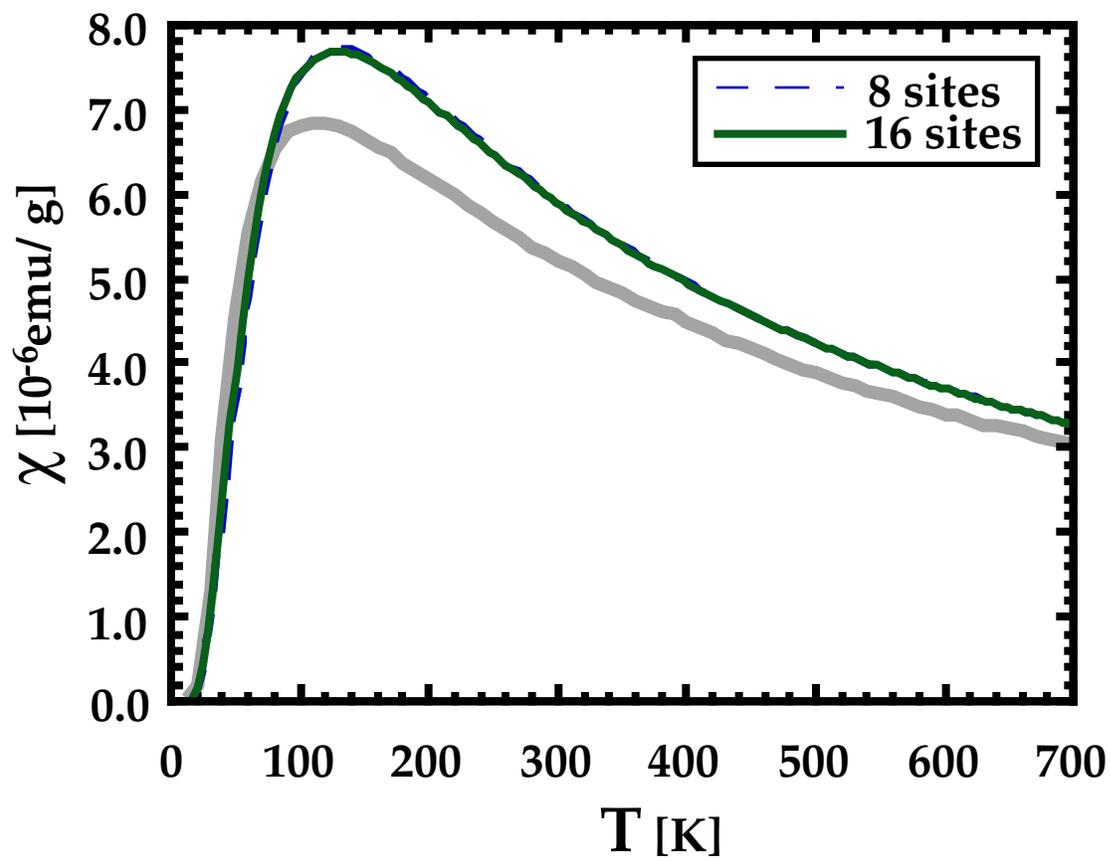
\end{document}